\documentclass[aps,12pt,onecolumn, superscriptaddress]{revtex4-2}
\usepackage{amssymb,graphicx,color,amsmath,xfrac,bm,array,oldstyle,setspace, soul}

\usepackage[mathletters]{ucs}
\usepackage[utf8x]{inputenc}

\definecolor{lightblue}{rgb}{0.17,0.39,1}

\usepackage[bookmarks, colorlinks=true, breaklinks, pdftitle={RuCl3  SI}, pdfauthor={Kim,Brad,Ross,Arkady}]{hyperref}
\hypersetup{linkcolor=lightblue,citecolor=lightblue,filecolor=black,urlcolor=lightblue}

\newcommand{\para}[1]{\left(#1\right)}
\newcommand{\degm}{{\ensuremath{^∘}}}
\renewcommand{\deg}{{$\degm$}}
\newcommand{\av}[1]{\ensuremath{\left\langle{\!{#1}\!}\right\rangle}}

\newcommand{\matr}[1]{\para{\begin{matrix}#1\end{matrix}}}

\newcommand{\sqmatr}[1]{\para{\begin{array}{l@{\hspace{2em}}l@{\hspace{2em}}l} #1 \end{array}}}

\newcommand{\B}{ {\ensuremath{\bm{B}}}}

\newcommand{\g}{ {\ensuremath{\hat{g}}}}

\newcommand{\rucl}{RuCl$_3$}

\begin{document}
\title{Scale-invariant magnetic anisotropy in RuCl$_3$ at high magnetic fields} 
\author{K. A. Modic}
\email[Email: ]{kmodic@ist.ac.at}
\affiliation{Institute of Science and Technology Austria, Am Campus 1, 3400 Klosterneuburg, Austria}
\affiliation{Max-Planck-Institute for Chemical Physics of Solids, Noethnitzer Strasse 40, D-01187 Dresden, Germany}
\author{Ross D. McDonald}
\affiliation{Los Alamos National Laboratory, Los Alamos, NM 87545, USA}
\author{J. P. C. Ruff}
\affiliation{Cornell High Energy Synchrotron Source, Cornell University, Ithaca, NY 14853, USA}
\author{Maja D. Bachmann}
\affiliation{Max-Planck-Institute for Chemical Physics of Solids, Noethnitzer Strasse 40, D-01187 Dresden, Germany}
\affiliation{Department of Applied Physics and Geballe Laboratory for Advanced Materials, Stanford University, Stanford, CA 94305, USA}
\author{You Lai}
\affiliation{Los Alamos National Laboratory, Los Alamos, NM 87545, USA}
\affiliation{Florida State University, Tallahassee, FL 32310, USA}
\affiliation{National High Magnetic Field Laboratory, Florida State University, Tallahassee, FL 32310, USA}
\author{Johanna C. Palmstrom}
\affiliation{Department of Applied Physics and Geballe Laboratory for Advanced Materials, Stanford University, Stanford, CA 94305, USA}
\author{David Graf}
\affiliation{National High Magnetic Field Laboratory, Florida State University, Tallahassee, FL 32310, USA}
\author{Mun Chan}
\affiliation{Los Alamos National Laboratory, Los Alamos, NM 87545, USA}
\author{F. F. Balakirev}
\affiliation{Los Alamos National Laboratory, Los Alamos, NM 87545, USA}
\author{J. B. Betts}
\affiliation{Los Alamos National Laboratory, Los Alamos, NM 87545, USA}
\author{G. S. Boebinger}
\affiliation{Florida State University, Tallahassee, FL 32310, USA}
\affiliation{National High Magnetic Field Laboratory, Florida State University, Tallahassee, FL 32310, USA}
\author{Marcus Schmidt}
\affiliation{Max-Planck-Institute for Chemical Physics of Solids, Noethnitzer Strasse 40, D-01187 Dresden, Germany}
\author{D. A. Sokolov}
\affiliation{Max-Planck-Institute for Chemical Physics of Solids, Noethnitzer Strasse 40, D-01187 Dresden, Germany}
\author{Philip J. W. Moll}
\affiliation{Max-Planck-Institute for Chemical Physics of Solids, Noethnitzer Strasse 40, D-01187 Dresden, Germany}
\affiliation{Institute of Material Science and Engineering, École Polytechnique Fédéral de Lausanne (EPFL), 1015 Lausanne, Switzerland}
\author{B. J. Ramshaw}
\affiliation{Laboratory for Atomic and Solid State Physics, Cornell University, Ithaca, NY 14853, USA}
\author{Arkady Shekhter}
\affiliation{National High Magnetic Field Laboratory, Florida State University, Tallahassee, FL 32310, USA}

\pacs{...}
\date{\today }
\maketitle

\setlength{\parindent}{0.cm}
\setlength{\parskip}{0.8cm}
 
{\bf 
In \rucl, inelastic neutron scattering and Raman spectroscopy reveal a continuum of non-spin-wave excitations that persists to high temperature, suggesting the presence of a spin liquid state on a honeycomb lattice. In the context of the Kitaev model, magnetic fields introduce finite interactions between the elementary excitations, and   thus the effects of high magnetic fields -- comparable to the spin exchange energy scale -- must be explored. Here we report measurements of the magnetotropic coefficient -- the second derivative of the free energy with respect to magnetic field orientation -- over a wide range of magnetic fields and temperatures. We find that magnetic field and temperature compete to determine the magnetic response in a way that is independent of the large intrinsic exchange interaction energy. This emergent scale-invariant magnetic anisotropy provides evidence for a high degree of exchange frustration that favors the formation of a spin liquid state in RuCl$_3$.
}

The Kitaev model \cite{Kitaev2006} has directed the search for quantum spin liquids towards honeycomb networks of transition-metal ions where the exchange interaction is mediated via edge-shared octahedra \cite{Kasahara2018, Kasahara2018b, Sears2017, Wolter2017, Sandilands2015, Banerjee2016, Banerjee2017, Baek2017, Wang2017, Ponomaryov2017}. These systems have spin-anisotropic exchange interactions \cite{Jackeli2009, Savary2016,Modic2014, Chun2015}, and consequently may host a spin liquid ground state. The recent discovery of fractionalized excitations in \rucl\ has provided compelling evidence that the Kitaev model can be studied in a real system \cite{Banerjee2016, Banerjee2017, Sandilands2015, Kasahara2018b}. 
 
Kitaev’s model has only one intrinsic energy scale -- the spin-anisotropic exchange interaction $J_\text{K}$. Therefore in the pure Kitaev model, one should not expect significant changes in its behavior until the temperature is of order the exchange interaction energy $ J_\text{K} \sim$ 100 K \cite{Majumder2015, Banerjee2016}. In RuCl$_3$, the continuum of excitations observed in Raman and neutron scattering experiments survives up to a temperature scale that is comparable to this anisotropic exchange energy \cite{Banerjee2017,Sandilands2015}. However, it is known that the exchange interactions in RuCl$_3$ are not purely Kitaev-like, and it becomes unstable due to additional interactions. Indeed, all candidate materials ((Na,Li)$_2$IrO$_3$ and \rucl) are prone to antiferromagnetic order in the low-field and low-temperature part of the phase diagram \cite{Modic2014, Kimchi2014, Chaloupka2013, Das:2019}. 
 These ordered magnetic states are easily suppressed at temperatures and magnetic fields well below the exchange interaction energy scale \cite{Johnson2015, Kubota2015, Cao2016, Majumder2015, Modic2014, Modic2017, Leahy2017}, and recent theoretical studies have explored the possibility of a Kitaev-like spin liquid persisting to high magnetic fields \cite{Yoshitake:2020, Gordon:2019}. This calls for studies of Kitaev-like systems over a broad range of temperature and magnetic field where the underlying Hamiltonian can be studied directly, in the absence of magnetic order. 
 
 \begin{figure}[h!]
\centering
\includegraphics[width=1\linewidth, trim=0cm 20cm 0cm 0cm, clip=true]{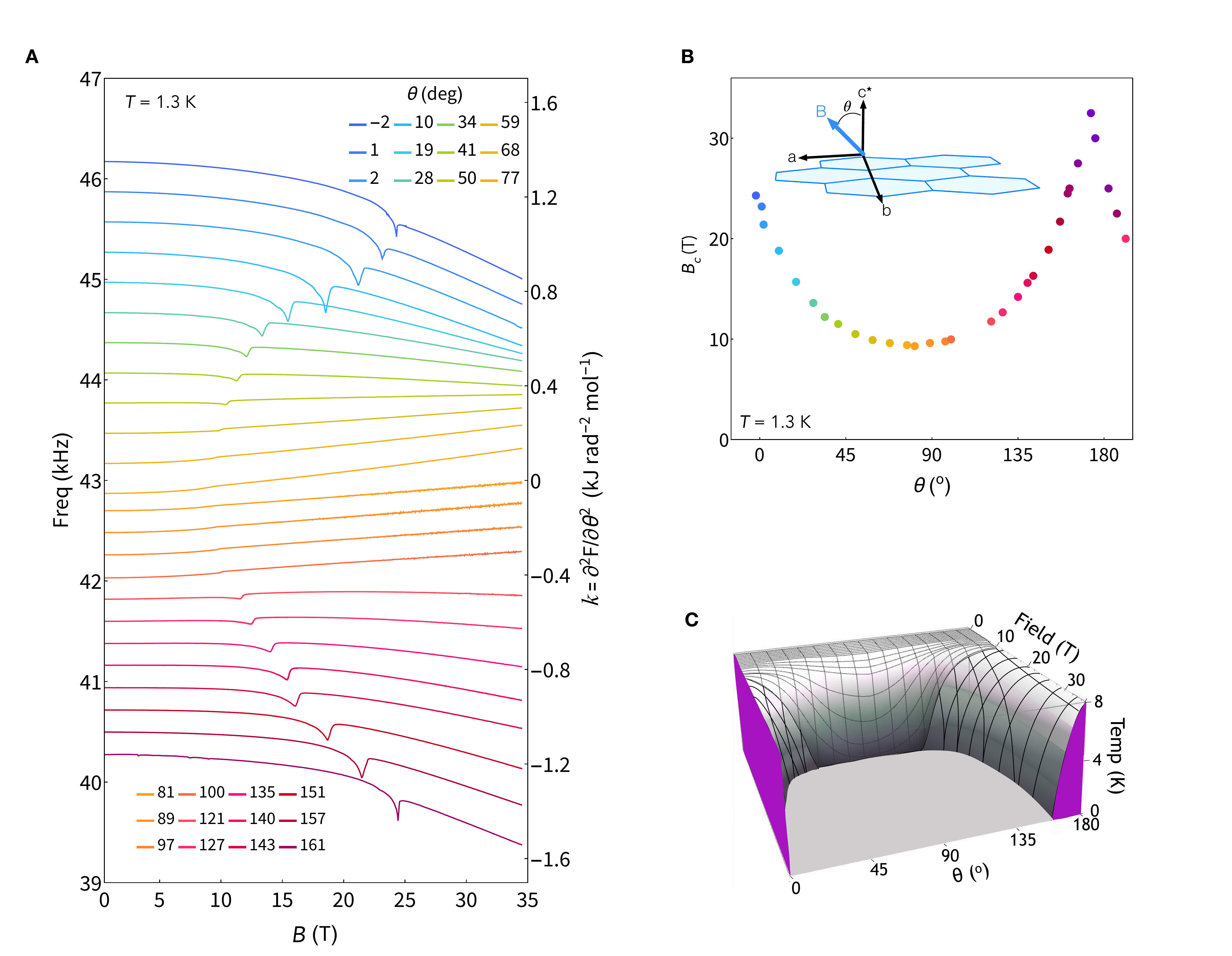}
\rule{27em}{0.5pt}
\singlespace
\caption[Thermodynamics]{ A) Field scans of the magnetotropic coefficient $k$ at T = 1.3 K for several field orientations. $θ$ is defined as the angle between the applied magnetic field and the direction nearly perpendicular to the honeycomb planes (\autoref{sec:monoclinic}) and the caption of Figure 4 for details concerning the precise definition of $θ$). The in-plane field orientation is unknown for these measurements. For $θ \approx$ 90$^\circ$, the magnetotropic coefficient monotonically increases up to 34.5 T. The maximum frequency shift (at all fields) occurs when magnetic field lies in the honeycomb plane, suggesting a magnetically easy axis (see also Figure 4A). As field is rotated away from the plane, a sharp jump in the magnetotropic coefficient is observed as the second-order AFM phase boundary is crossed. B) The AFM transition field as a function of angle at T = 1.3 K, as determined by the data in panel A and \autoref{sec:adata}. The shift of the maximum and minimum in the transition field away from 0$^\circ$ and 90$^\circ$, respectively, is a result of the monoclinic crystal structure \cite{Johnson2015}, which allows the principal magnetic axes to freely rotate with magnetic field and temperature (\autoref{sec:monoclinic}). C) The anisotropic AFM phase boundary constructed from jumps observed in resonant torsion measurements (panel A, Figure 4A, and \autoref{sec:adata}).}
\label{fig:figure1}
\end{figure}
 
We use resonant torsion magnetometry to study the magnetic anisotropy of \rucl\ \cite{Gammel1988, Bishop-RSI-1985, Modic2018}. This technique measures the magnetotropic coefficient $k = ∂^2F/∂θ^2$, where $F$ is the free energy and $θ$ is the angle of rotation of the magnetic field in the crystal. The magnetotropic coefficient represents a material's magnetic ``rigidity" with respect to rotation in a magnetic field. It is detected as a shift in the natural frequency of a freely suspended cantilever onto which the sample is mounted \cite{Modic2018}. \rucl\ orders antiferromagnetically below $T_N ≈ $ 7 K \cite{Cao2016}. However, multiple magnetic transitions were observed in the vicinity of the AFM phase boundary, which were shown to be due to additional domains of AB stacking or multiple monoclinic domains \cite{Cao2016, Kubota2015, Majumder2015}. The high sensitivity of our technique to magnetic anisotropy allows us to overcome the propensity for multiple domains in large samples by measuring $\sim$10-100 ng crystals of \rucl\ (roughly $10^5$ times smaller than those used in other techniques). Our measurements reveal a single magnetic phase boundary (\autoref{fig:figure1}), with no indication of other phase transitions at fields above 10 T.
 
We first map out the AFM phase boundary to identify the field orientation where the unordered state can be accessed over the broadest field range. \autoref{fig:figure1} shows the magnetic field evolution of the magnetotropic coefficient of \rucl\ at 1.3 K. A sharp jump in the magnetotropic coefficient outlines the anisotropic boundary of the AFM phase in the magnetic field magnitude, field orientation, and temperature phase space. This jump, both in magnetic field scans (\autoref{fig:figure1}A) and crystal rotation scans (\autoref{fig:figure4}A), is always {\emph{down}} upon entry into the ordered state as required for thermodynamic coefficients by Le Chatelier's principle \cite{Modic2018, Callen1985} (supplementary information \autoref{sec:thermodynamics}). \autoref{fig:figure1}C shows a schematic mapping of the AFM phase boundary constructed from additional data taken at multiple temperatures (\autoref{sec:adata}). The AFM transition field $B_c$ reaches a minimum of $\sim$10 T when the field is applied in the honeycomb plane ($θ$ = 90$^\circ$ in \autoref{fig:figure1}), and has a weak in-plane field orientation dependence \cite{Lampen2018}. $B_c$ increases beyond 30 T as the magnetic field is rotated toward the direction perpendicular to the honeycomb planes (\autoref{fig:figure1}B). 

\begin{figure}[h!]
\centering
\includegraphics[width=1\linewidth, trim=0cm 15cm 0cm 7cm, clip=true]{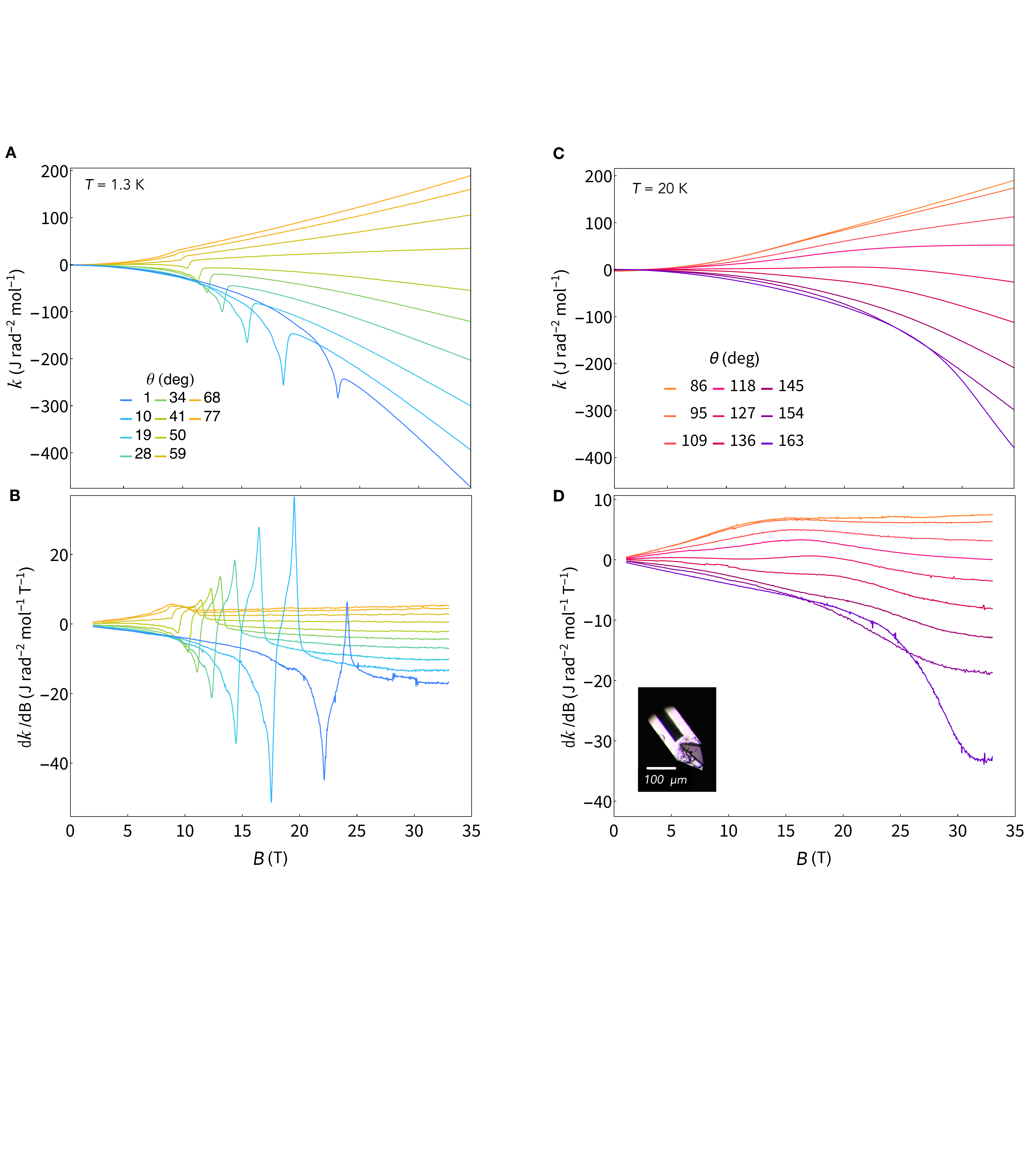}
\rule{27em}{0.5pt}
\singlespace
\caption[Field dependences]{ A) Angle dependence of the magnetotropic coefficient $k$ as a function of magnetic field at T = 1.3 K -- at low field the system is in the AFM phase. Only some of the curves from \autoref{fig:figure1}A are shown for clarity. B) The corresponding field-derivatives of $k$ show saturation once the AFM state is suppressed with magnetic field. C) Similar to the data shown in A, but at a temperature of 20 K---well outside the AFM state. D) The corresponding field-derivatives of the data shown in C, also showing saturation at high magnetic fields. The inset shows the sample mounted on the lever.}
\label{fig:figure2}
\end{figure}
 
For a wide range of field orientations near the honeycomb planes (90$^\circ$ in \autoref{fig:figure1}A), we observe a linear-in-field dependence of the magnetotropic coefficient above $B_c$. Therefore, we focus on a subset of angles near the $ab$-plane, where $B_c$ is minimal, in order to study this anomalous behavior in the broadest field range (\autoref{fig:figure2}). We note that unlike torque \cite{Leahy2017}, the magnetotropic coefficient does not vanish near the crystallographic directions, allowing direct measurement of magnetic anisotropy along these directions \cite{Modic2018}. Because $k$ is the second angular derivative of the free energy, the linear-in-field dependence of $k$ (\autoref{fig:figure2}A \&\ \autoref{fig:figure2}C) requires a linear-in-field dependence of the anisotropic part of the free energy itself \cite{Modic2018}. Put differently, the field dependence of $k$ captures the anisotropic part of the field-dependent free energy. The anomalous high-field behavior of the free energy in \rucl\ can be seen most clearly in the saturation of the field-derivatives of $k$ at 1.3 K (\autoref{fig:figure2}B). This saturation at high fields persists to at least 40 K -- well above the N\'{e}el temperature (\autoref{sec:adata}), indicating that this behavior is not associated with AFM order. The field at which a crossover from quadratic- to linear-in-field behavior occurs is pushed to higher fields with increasing temperature. This indicates a competition between the energy scales associated with temperature and magnetic field that is already evident in the comparison between the 20 and 40 K data (\autoref{fig:figure2}D \&\ \autoref{sec:adata}). The fact that this competition is evident at such low fields and temperatures is surprising because of the large intrinsic exchange energy $J_\text{K} \sim$ 100 K \cite{Johnson2015, Banerjee2016, Sandilands2015}. 

\begin{figure}[h!]
\centering
\includegraphics[width=1\linewidth, trim=0cm 10cm 0cm 9cm, clip=true]{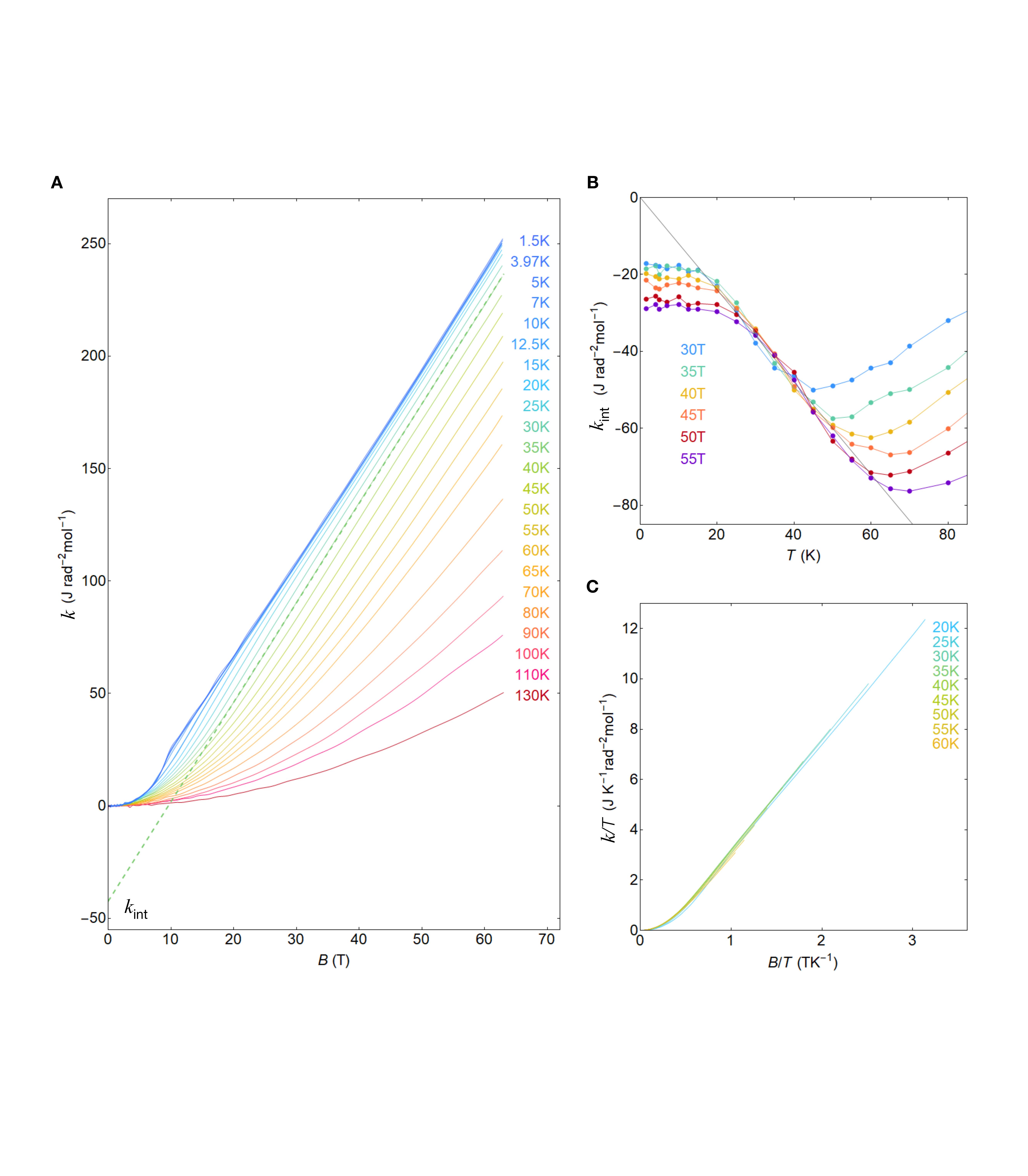}
\singlespace
\caption[]{ A) Pulsed magnetic field measurements of the magneotropic coefficient up to 64 T, with field applied in the honeycomb planes. At low fields (B $≤$ 10 T), $k$ is quadratic in field and proportional to $(χ_\parallel-χ_\perp) B^2$, which was used to calibrate the absolute value of $k$ (\autoref{sec:cali}). At low temperatures, $k$ is linear-in-field above the AFM transition up to the highest measured field. As temperature is increased, the onset field of the linear behavior increases approximately linearly with temperature. For temperatures above 60 K, 64 T is not sufficient to reach the linear regime. B) The temperature dependence of the zero-field intercept $k_{int}$ defined in panel A. Each curve is obtained from the zero-field intercept of a linear fit in a 10 T field range around the field value indicated in the legend. In the intermediate temperature range, all curves collapse onto a straight line with zero intercept, indicating the temperature range over which scale-invariant magnetic anisotropy is observed. The saturation behavior at low temperature indicates the low-energy cutoff of the scale-invariance. The deviation from linearity at high temperature indicates the insufficient field range to reach the linear-in-field behavior. C) The magnetotropic coefficient normalized by temperature and plotted versus $B/T$ shows a collapse of all curves in the temperature range where scaling behavior is expected based on panel B.}
\label{fig:figure3}
\end{figure}

To explore the temperature-field competition, we performed a set of high-field measurements (\autoref{sec:pulsed}) over a broader range of temperatures, with fields up to 64 T applied in the honeycomb plane (\autoref{fig:figure3}). At low temperatures, the magnetotropic coefficient increases linearly with magnetic field over the entire field range once magnetic order is suppressed. At higher temperatures, the magnetotropic coefficient becomes linear in the high field limit (\autoref{fig:figure3}A). Despite the fact that there are three energy scales in this experiment -- temperature, field, and the exchange interaction energy -- it appears that the magnetic anisotropy is controlled by temperature and magnetic field alone. To investigate this quantitatively, we plot the intercept $k_{int}$, as defined in \autoref{fig:figure3}A. This intercept is linear in temperature between 20 and 60 K, shown in \autoref{fig:figure3}B. This linearity suggests scaling behavior of the form $k(T,B)/T = f(\mu_B B/k_BT)$ (\autoref{sec:scaling}). This scaling is demonstrated in \autoref{fig:figure3}C by the collapse of all $k/T$ vs $B/T$ curves over the temperature range where linearity occurs in \autoref{fig:figure3}B. Above 60 K, we have an insufficient field range to observe the scaling behavior. On the low temperature side, scaling is interrupted by the intrinsic energy scale of approximately 10 K possibly associated with AFM order. 


\begin{figure}[h!]
\centering
\includegraphics[width=.9\linewidth]{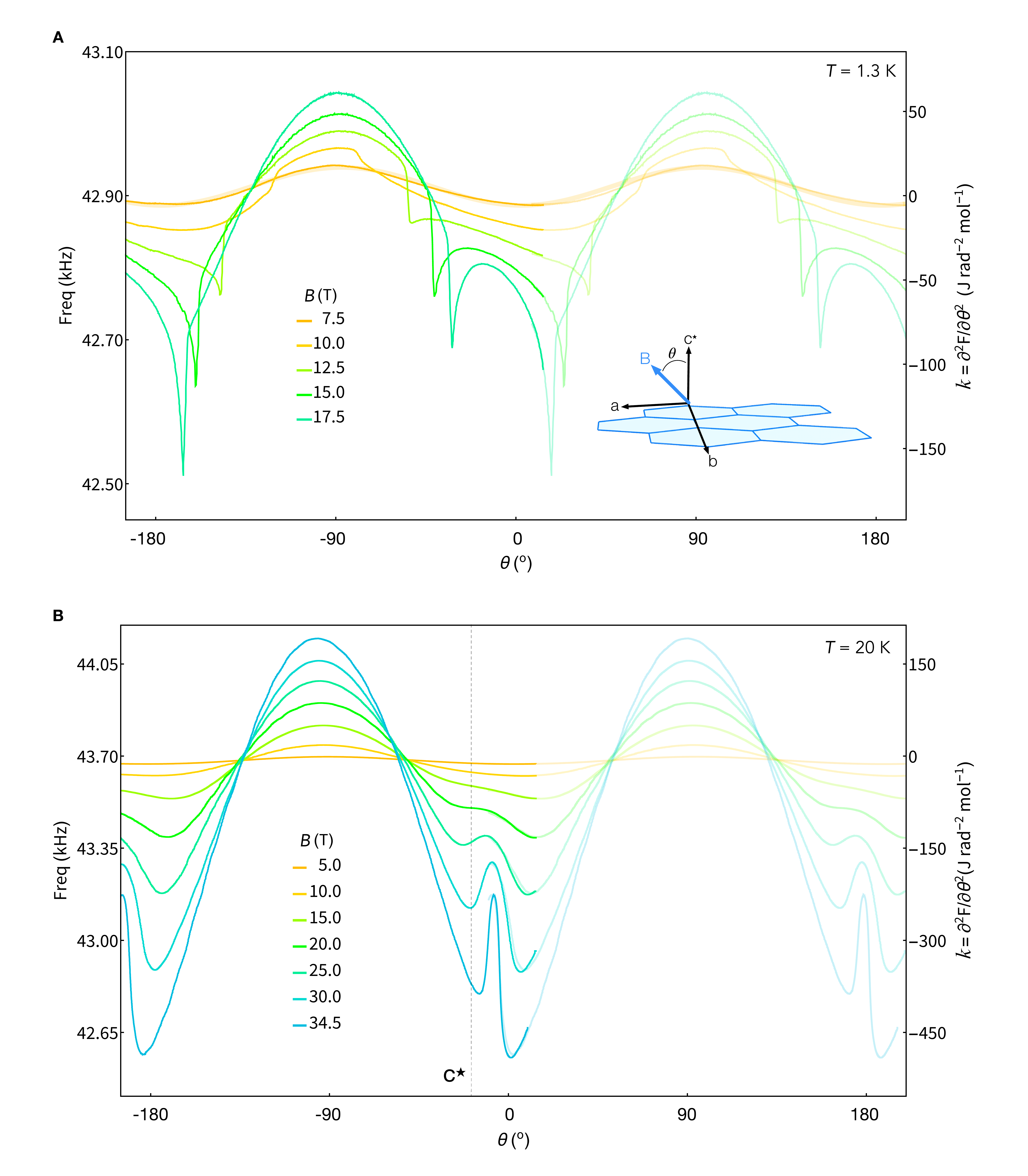}
\singlespace
\caption[]{ A) Angle scans of the magnetotropic coefficient, obtained by continuous rotation of the sample, in fixed magnetic fields ranging from 7.5 to 17.5 T. At 7.5 T, the magnetotropic coefficient (the angular derivative of torque \cite{Leahy2017}) follows the $\cos2θ$ angle dependence expected in the linear response regime $M_i = \chi_{ij}H_j$. The linear regime allows us to define the field orientations with respect to the principal directions of the magnetic susceptibility tensor. Additionally, this regime allows calibration between the measured frequency shift (left axis) and the absolute value of the magnetotropic coefficient (right axis, \autoref{sec:cali}). This is important because the absolute frequency at zero field depends on the elastic properties and the geometry of the cantilever, as well as the position of the sample on the cantilever. At higher magnetic fields, entry into and out of the AFM phase is marked by the sharp jump in $k$. Together with Figure 1A, these features outline the highly-anisotropic nature of the AFM phase (Figure 1C). B) The angle-dependent magnetotropic coefficient at 20 K -- outside of the AFM state. With increasing magnetic field, a spike develops for field near the $c^\star$ direction. Note that the position of the spike shifts with increasing magnetic field, which is due to the increasing importance of the exchange interactions within the honeycomb planes.}
\label{fig:figure4}
\end{figure}
 
The $B/T$ scaling of $k/T$ in high fields is in apparent contradiction to the angle dependence of the magnetotropic coefficient in \autoref{fig:figure4}. Panel A shows the angle dependence of $k$ at 1.3 K. At 7.5 T, as the crystal is rotated, the system is always in the AFM phase and the phase boundary is never crossed. In this regime, the magnetotropic coefficient follows a $\cos2θ$ dependence as expected \cite{Modic2018}. At higher fields, rotation in a fixed field causes the system to cross the AFM phase boundary, in accordance with \autoref{fig:figure1}C. At 20 K, the AFM phase is suppressed at all angles and magnetic fields, but anisotropy in $k$ as a function of angle is still observed. The strong deviation from $\cos2θ$ at higher fields reflects nonlinearity in the magnetization as a function of field. When the magnetic field crosses the direction perpendicular to the honeycomb planes ($c^{\star}$ in \autoref{fig:figure4}), $k$ exhibits a ``spike”, which becomes more prominent as the magnetic field is increased. This spike cannot be produced by an anisotropic $g$-factor alone \cite{Riedl:2019}, but requires exchange interactions that are of order the size of the magnetic field. This is demonstrated by both mean-field calculations (\autoref{sec:spike}) and exact diagonalization \cite{Riedl:2019}. The necessity for a relatively large exchange interaction (of order 60 K \cite{Riedl:2019}) sets up an apparent paradox with the data in \autoref{fig:figure3}, which suggests a small intrinsic energy scale no greater than 10 K. 

Scale-invariance is not foreign to spin systems. For example, Curie’s law for a gas of non-interacting spins is a direct consequence of the competition between the energy scales associated with magnetic field and temperature -- the only energy scales in the system \cite{Blundell}. This scale-invariance encapsulates the fact that the thermodynamic behavior is determined entirely by this competition of {\it external} energy scales. We find similar scale-invariance in the field and temperature dependence of the magnetotropic coefficient of RuCl$_3$. This is particularly surprising because the exchange interactions in RuCl$_3$ are thought to be quite large based on previous experiments \cite{Johnson2015, Banerjee2016, Sandilands2015} and the angle dependence of $k$ reported here (\autoref{sec:spike}). One possible mechanism for this ``emergent'' scale-invariance is local frustration of the exchange interactions. This can be as simple as competition between different exchange components (e.g. diagonal versus off-diagonal exchange), or it can be a result of frustration between purely anisotropic spins (e.g. the Kitaev interaction). The former is illustrated by our mean-field model: in general, $B/T$ scaling is destroyed by large exchange interactions, but can be restored when the magnitude of different exchange components are fine-tuned (e.g. $\Gamma$ and $J$ or $K$, see \autoref{sec:scaling}). This is a ``classical'' picture of magnetic frustration -- quantum systems can relieve this frustration by delocalizing their excitations, building long-range correlations between spins and favoring the development of a spin liquid state \cite{Anderson:1973}. These long-range correlations result in slower response times and lower characteristic energy scales. In this sense, the small effective exchange interaction and subsequent scale-invariance provide evidence for a spin liquid state in \rucl\ in a broad field and temperature range.

\section*{Acknowledgments}
 
The authors are grateful to Michael Baenitz, Ali Bangura, Radu Coldea, George Jackeli, Steve Kivelson, Steven Nagler, Roser Valenti, Chandra Varma, Stephen Winter, and Jan Zaanen for insightful discussions. Samples were grown at the Max Planck Institute for Chemical Physics of Solids. The DC-field measurements were done at the National High Magnetic Field Laboratory (NHMFL) in Tallahassee, FL. The pulsed-field measurements were done in the Pulsed Field Facility of the NHFML in Los Alamos, NM. X-ray analysis was done at Cornell University. Research conducted at the Cornell High Energy Synchrotron Source (CHESS) is supported by the National Science Foundation under award DMR-1332208. The DC- and pulsed-field work at the NHMFL is supported through the National Science Foundation Cooperative Agreement numbers DMR-1157490 and DMR-1644779, The United States Department of Energy, and the State of Florida. R.D.M. acknowledges support from LANL LDRD-DR 20160085 topology and strong correlations. Y.L. is supported by the US Department of Energy through the LANL/LDRD program and the G.T. Seaborg institute. J.C.P. is supported by a Gabilan Stanford Graduate Fellowship and a NSF Graduate Research Fellowship (grant DGE-114747). Sample characterization at Cornell was supported as part of the Institute for Quantum Matter, an Energy Frontier Research Center funded by the U.S. Department of Energy, Office of Science, Office of Basic Energy Sciences under Award Number DE-SC0019331.

 
 
 
\renewcommand{\thesection}{S\arabic{section}}
\renewcommand{\thefigure}{S\arabic{figure}}
\renewcommand{\theequation}{S\arabic{equation}}

\cleardoublepage
\section*{Supplementary Information}

\section{Thermodynamics of the AFM phase boundary}
\label{sec:thermodynamics}

The magnetotropic coefficient is a second derivative of the free energy and as such, it is a thermodynamic coefficient. Choosing the temperature and the angle of rotation as thermodynamic variables, the entropy and magnetic torque are defined as first derivatives: $S=-dF/dT$ and  $τ=dF/d\theta$. These thermodynamic coefficients can be written in a matrix, 
\begin{align}\label{eq:thermodynamiccoefficients}
\matr{ 
		dS \\ 	
		dτ
} = 
\matr{ 
		\frac{C}{T} & \para{ \frac{∂S}{d\theta}}_{\!T} \\
		\para{\frac{∂τ}{dT}}_{\!\!\theta}  & k 
} \matr{
		dT \\
		d\theta
},
\end{align}
where the off-diagonal terms are magnetocaloric coefficients. \autoref{eq:thermodynamiccoefficients} implies an Ehrenfest relation between the jump in heat capacity and the jump in the magnetotropic coefficient across a second order phase boundary, 
\begin{equation}
Δk = - ΔC/T_c \left(\frac{dT_c}{d\theta}\right)^{\!\!2}_{\!\!\!B},
\label{eq:ehrenfest}
\end{equation}
 where $\para{dT_c/d\theta}$ is the change in the transition temperature when the sample is rotated in a fixed magnetic field. Because $\Delta C$ is always positive, Eq.~{eq:ehrenfest} requires that the jump in the magnetotropic coefficient is always down as we enter into the broken-symmetry phase.

\begin{figure}[h!]
\centering
\includegraphics[width=.8\linewidth, trim=0cm 0cm 0cm 0cm, clip=true]{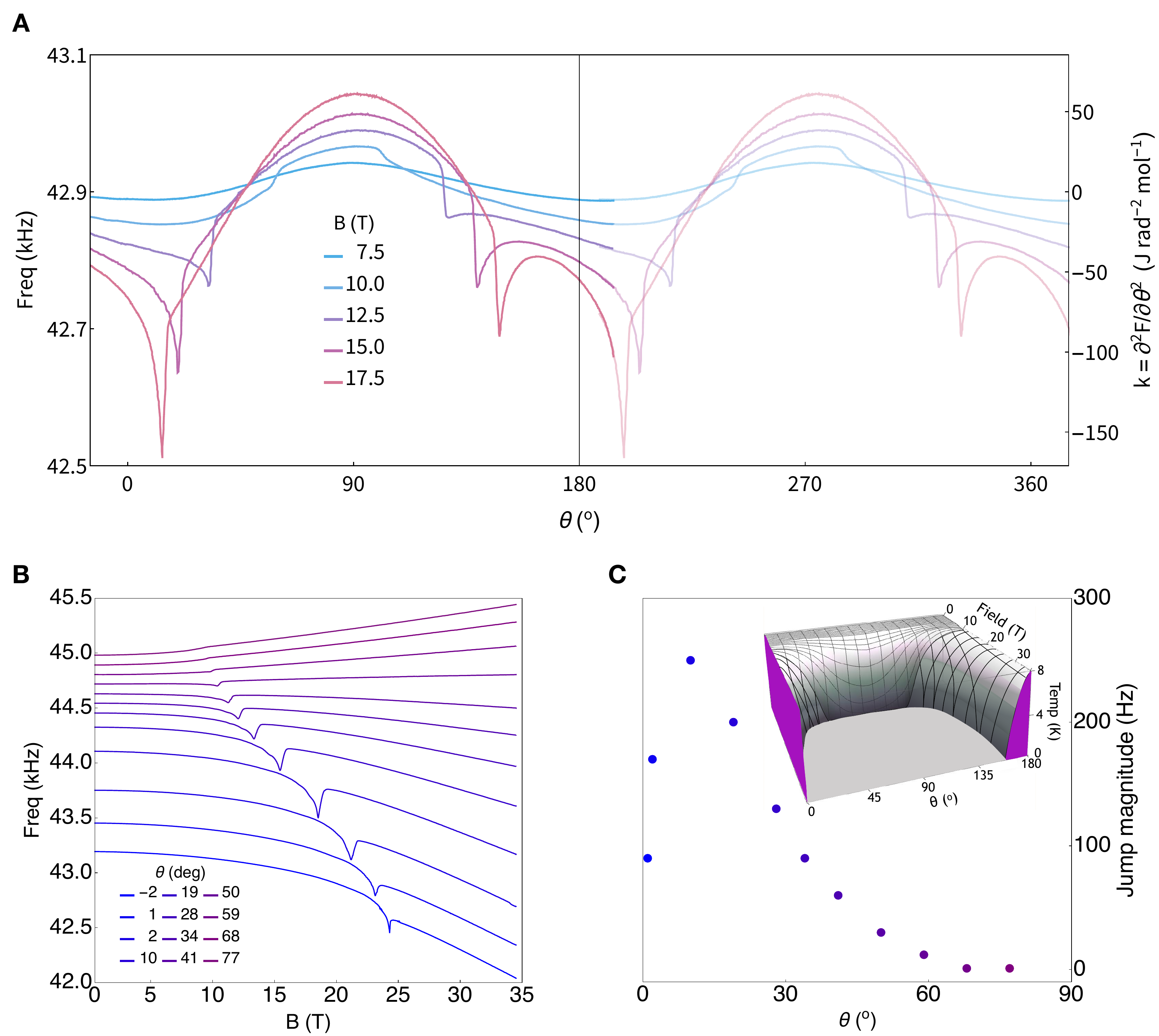}
\rule{27em}{0.5pt}
\singlespace
\caption{ A) Angle scans of the magnetotropic coefficient at 1.3 K, obtained by continuous rotation of the sample in fixed magnetic fields ranging from 7.5 to 17.5 T. At 7.5 T, the angle dependence of the magnetotropic coefficient is close to the $\cos2θ$ behavior expected in the linear response regime. Rotation of the crystal in magnetic fields $≥$10 T exhibits transitions into and out of the AFM phase. Entry into the ordered state is observed as a sharp jump down in the magnetotropic coefficient \cite{Modic2018}. Together with panel B, these features are used to map out the anisotropic AFM phase boundary (inset of C). B) Field scans of the magnetotropic coefficient at 1.3 K for several field orientations $θ$, where $θ$ is determined by a fit to the low-field $\cos2θ$ dependence. C) Angle dependence of the jump magnitude at the AFM boundary. As magnetic field is rotated away from the honeycomb planes at $θ = 90^∘$, the magnitude of the jump increases and reaches a maximum at $∼$10$^∘$.}
\label{fig:jump}
\end{figure}

\autoref{fig:jump}A shows the angle dependence of the magnetotropic coefficient at 1.3 K for fixed magnetic fields (the same data as in Figure 4A of the main text). \autoref{fig:jump}B shows a subset of the field scans from Figure 1A of the main text in order to emphasize the evolution of the jump magnitude (\autoref{fig:jump}C). The angle dependence of the jump indicates a stronger thermodynamic signature of the phase transition near the $c$-axis, with a maximum occurring at $∼$10$^∘$ away -- in agreement with the magnetically rotated principal direction of the susceptibility (\autoref{sec:monoclinic}). Note that the angle dependence is parametric: $Δk$ is a function on the surface of the AFM boundary $Δk(B_c,T_c,θ)$ (inset in panel C), where (at fixed $T_c$) $B_c$ itself is a function of $θ$. By symmetry, $Δk(B_c,T_c,θ)$ must vanish along the symmetry directions in the lattice where the derivative $(∂ T_c/∂ θ)_B$ on a fixed-field section of the AFM boundary vanishes. This is observed near the honeycomb planes (\autoref{fig:jump}C and inset).

\section{Rotation of the principal magnetic axes with temperature and magnetic field}
\label{sec:monoclinic}

\begin{figure}[h!]
\centering
\includegraphics[width=1\linewidth, trim=0cm 0cm 0cm 0cm, clip=true]{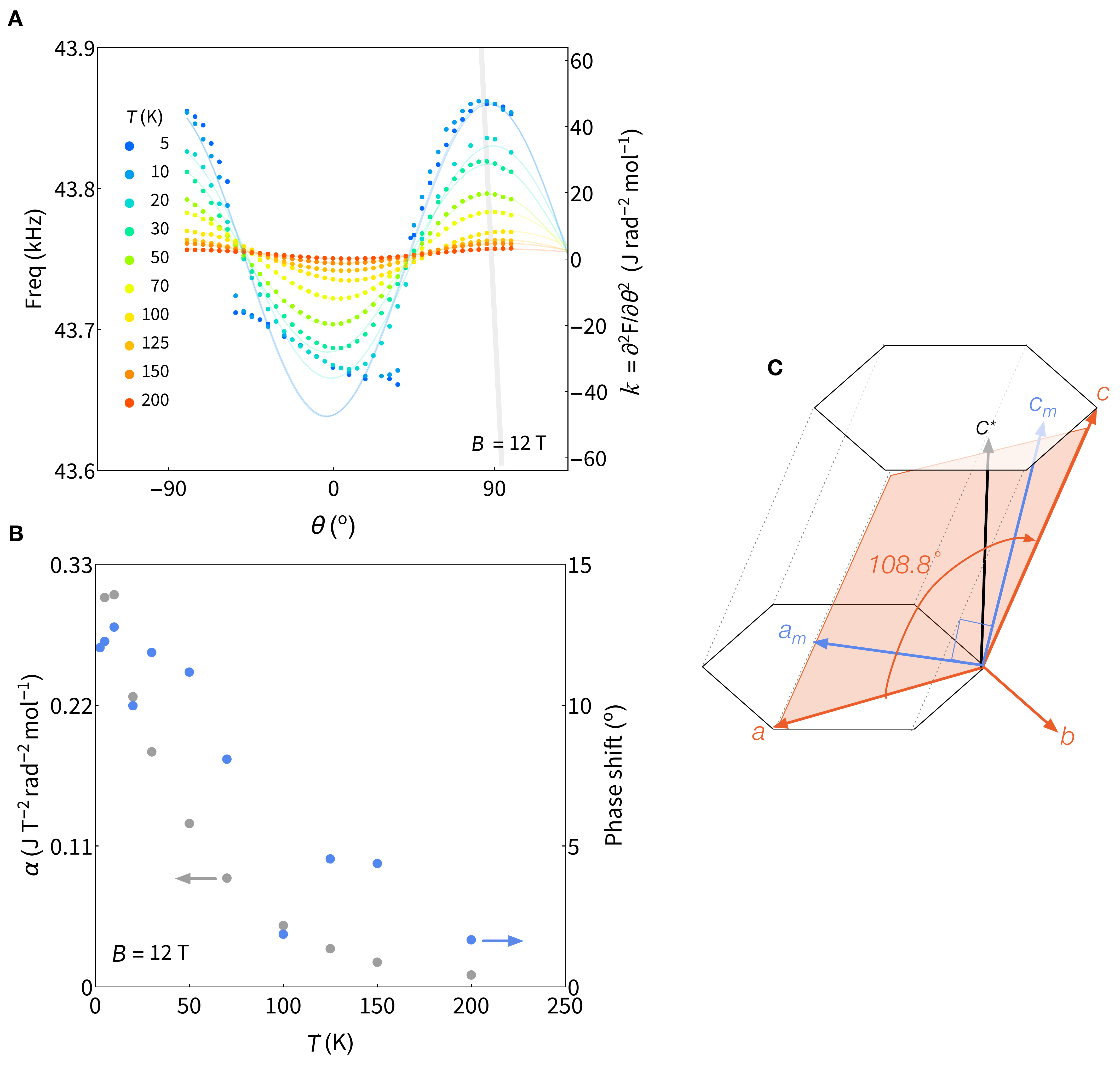}
\rule{27em}{0.5pt}
\singlespace
\caption[]{ A) The angle-dependent magnetotropic coefficient at 12 T for temperatures ranging from 5 to 200 K. At 12 T, the magnetotropic coefficient follows a $\cos2θ$ angle dependence at temperatures above the AFM phase. The amplitude of the $\cos2θ$ dependence is directly proportional to the magnetic anisotropy coefficient $α_{ij}=χ_i-χ_j$ in the plane of the vibrating lever (left axis in panel B). B) The magnetic anisotropy monotonically increases as the AFM phase is approached at lower temperatures. This is accompanied by a shift in the phase of the $\cos2θ$ angle dependence (right axis), which is a direct consequence of the fact that the principal magnetic axes can rotate due to the monoclinic crystal structure. C) Schematic representation of the crystallographic and principal magnetic axes ($a_m$ and $c_m$) in the monoclinic crystal structure of \rucl.}
\label{fig:monoclinic}
\end{figure}

The monoclinic crystal structure of \rucl does not require that the principal components of the susceptibility tensor coalign with the crystallographic axes or with the honeycomb planes. As such, we define $θ$ using the low-field response of the magnetotropic coefficient, where it has the same principle magnetic axes as the magnetic susceptibility tensor $χ_{ij}$. In this limit, the magnetotropic coefficient exhibits a $\cos2θ$ angle dependence (\autoref{fig:monoclinic}A) that allows us to define three perpendicular principal components of the susceptibility tensor, for example as $a_m$, $b$ and $c_m$ shown schematically in \autoref{fig:monoclinic}C. The monoclinic angle in \rucl is perpendicular to the $b$-axis, which allows the other two principal components ($a_m$ and $c_m$) to freely rotate in the $ac$-plane with temperature. In other words, in a monoclinic system, both the magnitudes and the directions of the principle components of $χ_{ij}$ evolve with temperature. This is observed as a continuous phase shift in the $\cos2θ$ angle dependence (gray line in \autoref{fig:monoclinic}A and right axis in \autoref{fig:monoclinic}B). The magnetic axis $b_m$ is required by symmetry to coincide with the crystallographic $b$-axis at all temperatures.

\begin{figure}[h!]
\centering
\includegraphics[width=1\linewidth, trim=0cm 0cm 0cm 0cm, clip=true]{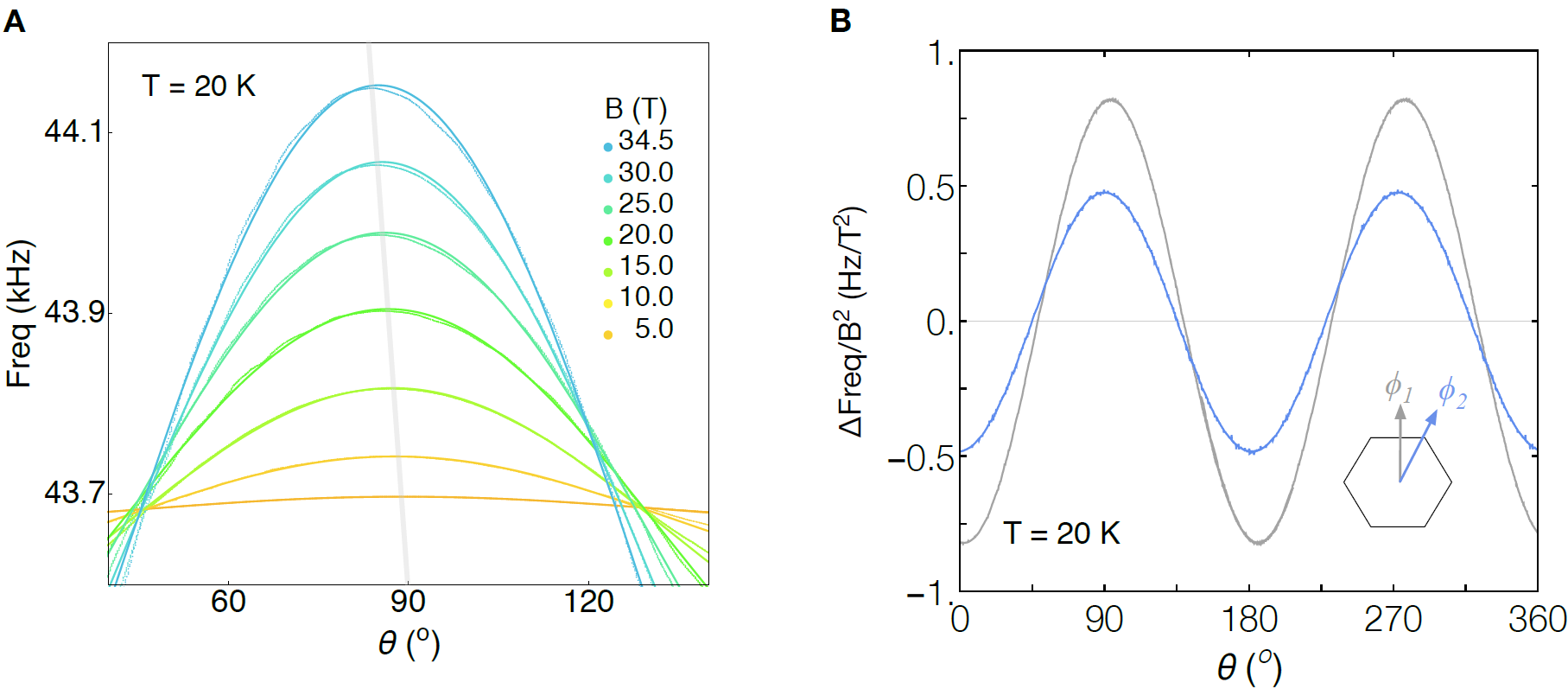}
\rule{27em}{0.5pt}
\singlespace
\caption[]{ A) A close-up of the data shown in Figure 4B of the main text to highlight the increasing role of the exchange interactions within the honeycomb plane with increasing magnetic field. B) The angle dependence of the magnetotropic coefficient normalized by magnetic field for rotation in the two azimuthal planes. Both the amplitude and phase of the $\cos2θ$ dependence depend on the $φ$ plane of rotation.}
\label{fig:monoclinic2}
\end{figure}

A similar realignment of the magnetic response occurs as we increase the magnetic field. Because the magnetic response is nonlinear at high fields, the angle dependence no longer follows a $\cos2θ$ dependence. Instead, one can follow the angle of the maximum in $k$ near the $ab$-plane, which shifts with increasing field (illustrated by the gray line in \autoref{fig:monoclinic2}A). This behavior indicates that the magnetic response at high fields is determined entirely by the exchange interactions within the honeycomb planes (i.e., the maximum $k$ for each curve approaches the $ab$-plane as magnetic field increases). Simultaneously, the minimum in the magnetotropic coefficient approaches the $c^⋆$-axis. The observed phase shift of $∼$10$^∘$ is qualitatively consistent with the reported monoclinic angle $β = 108.8^∘$ \cite{Cao2016} (i.e., the low field direction of $c_m$ is about halfway between the crystallographic $c$-axis and the direction $c^⋆$ perpendicular to the $ab$-plane).

One would expect that both the amplitude of the magnetic anisotropy and its orientation with respect to the crystal axes to depend on the $φ$ plane of rotation, even in the linear response regime. This is confirmed in \autoref{fig:monoclinic2}B for the magnetotropic coefficient normalized by magnetic field for two different planes of rotation. A change in amplitude and a slight shift in phase are observed because the principal magnetic axes map onto each plane of rotation.

\section{Resonant torsion magnetometry in pulsed magnetic fields}
\label{sec:pulsed}

\begin{figure}[h!]
\centerline{\includegraphics[width=1\linewidth, trim=0cm 0cm 0cm 0cm, clip=true]{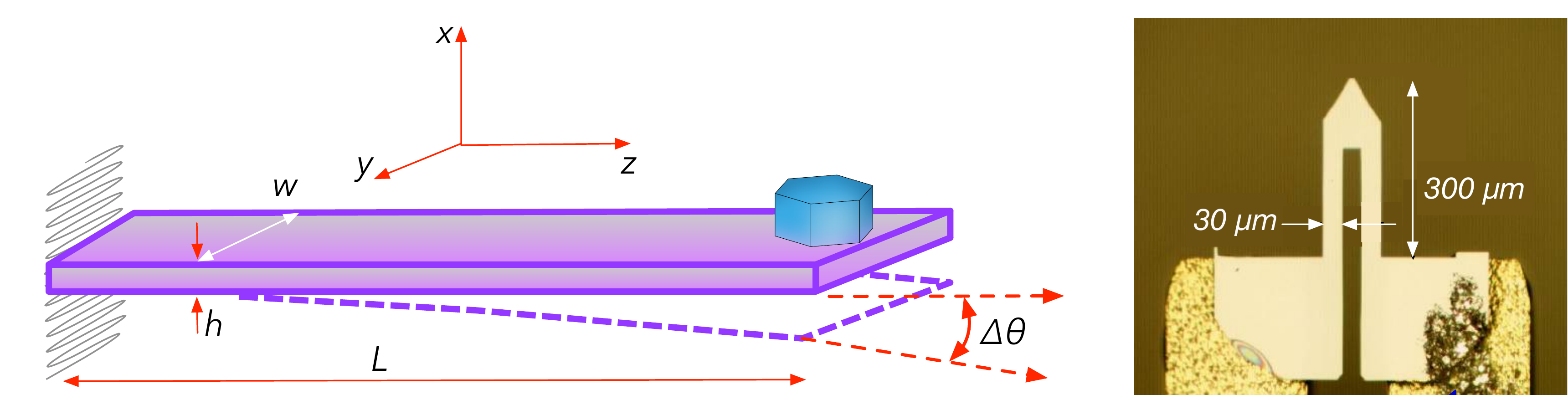}}
\singlespace
\caption{ Schematic of a thin bending lever (left) and the Akiyama silicon lever (right) \cite{Akiyama2010}. The thickness of the silicon tip is $h=3.7\;μ$m. The dashed wall on the left represents the rigid attachment to the transducer fork.}
\label{fig:lever}
\end{figure}

Resonant torsion measurements of the magnetotropic coefficient $k=d^2F/dθ^2$ of the sample rely on the lever being a simple harmonic oscillator,
\begin{align}\label{eq:effective0}
  E= \frac{I}2 \left(\frac{d∆θ}{dt}\right)^2 + \frac{K}2 ∆θ^2  \,, \qquad ω_0^2 = K/I\,.
\end{align}
Here, $∆θ(t)$ is the angle of rotation at the tip of the lever where the sample is attached (\autoref{fig:lever}). 

The resonant frequency $ω_0^2 = K/I$ of the lever is determined by the bending stiffness $K$, and moment of inertia $I$ \cite{Modic2018}. A small magnetically anisotropic sample mounted on the tip of the lever introduces an additional energy term, 
\begin{align}
\delta E_{\text{sample}} =  \frac{k}2 ∆θ^2\,,	
\end{align}
where $k$ is the magnetotropic coefficient of the sample. This additional energy term is responsible for a small frequency shift \cite{Modic2018},
\begin{align}\label{eq:freqshift}
(ω_0+Δω)^2 = \frac{k+K}{I}\,, \qquad 
\frac{Δω}{ω_0} ≈ \frac{k}{2K}.
\end{align}
In the resonant torsion magnetometry measurements, the magnetotropic coefficient is inferred from the resonant frequency shift via \autoref{eq:freqshift}. 

\begin{figure}[h!]
\centering
\includegraphics[width=0.9\linewidth]{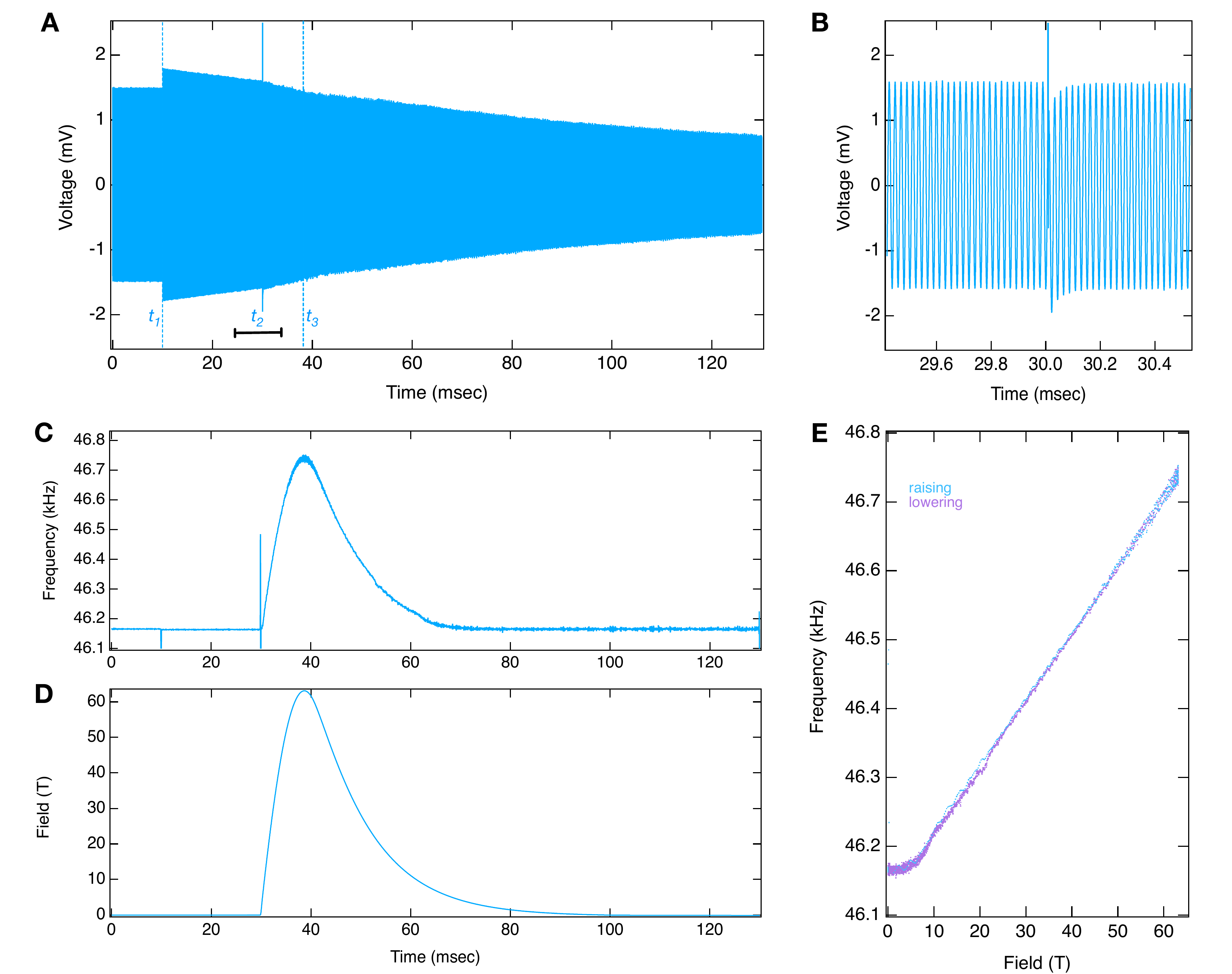}
\singlespace
\caption{ The magnetotropic coefficient of \rucl in pulsed fields. A) A piezoelectric-induced voltage during free oscillation of the lever. The lever is driven close to the resonant frequency to achieve a large initial amplitude. The piezoelectric voltage is about 1.5 mV at 100 mV drive and it corresponds to a bending oscillation of about 0.5$^{∘}$. The drive voltage is turned off at time $t_1=10$ msec (shortly after the start of digital recording). At $t_1$, the amplitude of the pickup voltage is increasing by about 20\%. This is because when the drive is turned on, the pickup voltage is a superposition of the piezo voltage and the voltage induced in the wires leading to the lever. Their relative magnitude and phase depends on wiring details and differences in the drive frequency and the natural frequency of the lever. At $t>t_1$, the measured pickup voltage is entirely determined by the piezo voltage on the stressed (by the lever) transducer fork. The magnet pulses 20 msec later, at time $t_2=30$ msec. Peak field is reached approximately 8 msec after that at time $t_3=38$ msec. B) Zoom-in of the region indicated by the black line in panel A, highlighting the start of the magnet pulse. C) Frequency of the lever oscillations over the duration of the entire pulse. The frequency is obtained by analyzing the Fourier transform of the pickup voltage with a sliding window of 200 μsec --corresonding to about 10 periods of oscillation. D). Time-evolution of the magnetic field in the multi-shot 65 T magnet system at the NHMFL-pulsed field facility in Los Alamos. E) Field-dependence of the frequency of oscillations obtained by plotting the data in panel C vs the data in panel D. The blue and purple curves show the rising and falling side of the field pulse, respectively.}
\label{fig:pulse}
\end{figure}

In DC magnetic fields, the frequency shift can be measured with a lock-in amplifier and tracked with a phase-locked loop (PLL), where the driving frequency is adjusted with negative feedback from the phase of the pickup \cite{Gammel1988, Bishop-RSI-1985, Modic2018}. Our existing implementation of the PLL \cite{custom-PLL} is not fast enough to reliably follow the resonant frequency in pulsed fields, where it shifts by 100's of Hz over a $∼$10 msec duration of the pulse.

To achieve the necessary resolution of the frequency shift in pulsed field measurements, we analyze the \emph{free} oscillation of the lever. We ring up the oscillation amplitude by driving the lever close to the resonant frequency and then turn off the driving voltage right before the magnet pulses (\autoref{fig:pulse}). The lever oscillates freely throughout the whole duration of the magnet pulse. In the adiabatic limit, the frequency shift closely reflects the static natural oscillation frequency at an instantaneous value of the magnetic field. As the pulse progresses, the energy dissipation in the lever, sample, and surrounding exchange gas reduce the amplitude of the lever oscillations.

The free evolution of the lever in the magnetic field is governed by
\begin{align}
 \frac{d^2∆θ}{dt^2} +ω_0(t)^2∆θ =0,
\end{align}
where the time dependence of the frequency $ω_0(t)^2 = ({K+k(t)})/{I}$ is a result of the field dependence of the magnetotropic coefficient $k(t) = k(B(t))$ during the pulse $B(t)$. We can check whether we are in the adiabatic limit by calculating the parameter $α$, defined as the relative change of frequency during one period of oscillation, $α = d(ω_0^{-1})/dt$. In our measurement (Figure 3 of the main text), the frequency $ω_0(t)$ shifts by about 500 Hz around 50 kHz when the field is pulsed to 65 T over 50 msec (on the falling side of the pulse), therefore $α ≈ 5×10^{-6}≪1$ and the measurements in Figure 3 of the main text are well within the adiabatic regime.

\section{Calibration of the magnetotropic coefficient}
\label{sec:cali}

\begin{figure}[h!]
\centering
\includegraphics[width=0.95\linewidth]{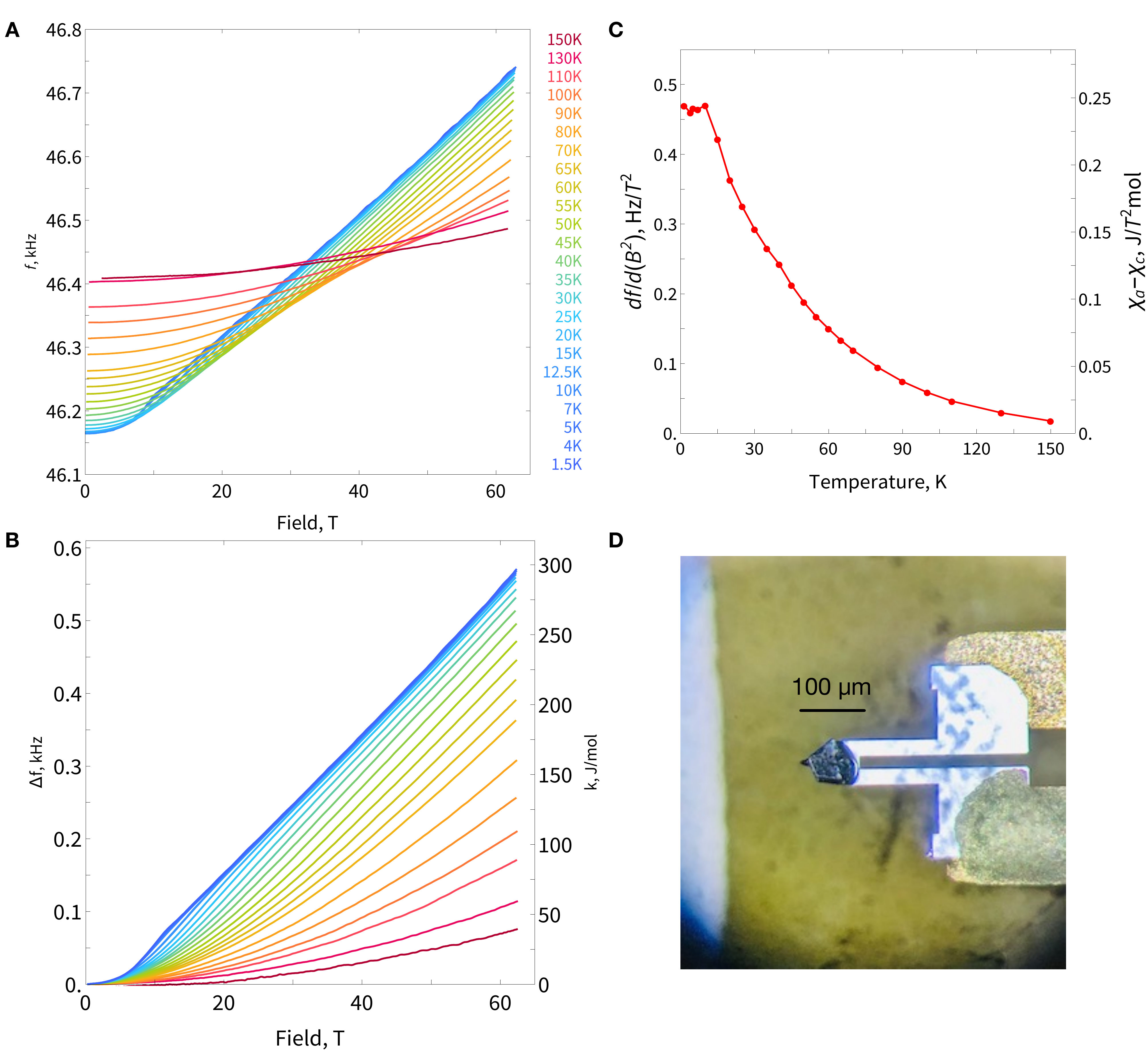}
\singlespace
\caption{
Resonant torsion measurements up to 63 T for temperatures ranging from 1.5 to 150 K.  A) The zero-field frequency shift is a result of small ($≲1\%$) changes in the elastic stiffness of the lever over the broad temperature range. B) The absolute frequency shift with respect to the zero-field frequency $Δf(B=0) =0$. The vertical axis on the right shows corresponding values of the magnetotropic coefficient (\autoref{sec:cali}). C) The low-field (linear regime) coefficient $a$ of frequency shift vs $B^2$, $Δ f = aB^2$. For the (fixed) orientation of the lever in this measurement $Δ f = (χ_1-χ_2)B^2$ where $χ_{1,2}$ are the principal components of the magnetic susceptibility. The vertical axis on the right shows the corresponding values of the magnetic anisotropy $χ_1-χ_2$ in per-mol units \cite{Banerjee2017}. D) Sample mounted on the lever for field aligned perpendicular to the honeycomb plane.}
\label{fig:60T}
\end{figure}

To obtain quantitative values for the magnetotropic coefficient (per unit volume, per mol, etc.), one must know the mass of the sample. In this work, the sample masses are all in the range of 10 - 100 ng (\autoref{fig:60T}D) for which we could not obtain a reliable mass measurement. Instead, we use the known magnetic susceptibility (per unit volume) of \rucl  \cite{Banerjee2016, Kubota2015, Baek2017, Sears2017}. For small fields (in the linear regime), the magnetotropic coefficient is related to the magnetic susceptibility via $k = B^2(χ_1-χ_2)\cos2θ$ where $χ_{1,2}$ are the two components of the magnetic susceptibility tensor along the two principal directions in the plane of rotation \cite{Modic2018}. 

The data in Figure 3 of the main text has been calibrated as follows. \autoref{fig:60T}C shows the coefficient of proportionality of the frequency shift vs $B^2$ in the low-field limit, which is proportional to $χ_1-χ_2$ for rotation out of the plane (when $\cos2θ=1$). At temperatures below 5 K, the magnetic anisotropy $χ_1-χ_2$ saturates at $24 ×10^{-3}$ emu/mol = 0.24 J/mol/T$^2$ \cite{Banerjee2016, Kubota2015, Baek2017, Sears2017}.  The measured value of $df/dB^2$ is 0.48 Hz/T$^2$ and is equivalent to 0.24 J/mol/T$^2$ or 1 Hz $↔$ 0.5 J/mol/rad$^2$ for the data in Figure 3 and \autoref{fig:60T} (the conversion shown on the right vertical axes in \autoref{fig:60T}B and \autoref{fig:60T}C). Similarly, this procedure was also used to calibrate the measurements in Figures 1, 2, and 4 of the main text, where $df/dB^2$ is 0.46 Hz/T$^2$ at an angle of $\theta = 100\degm$, which yields 1 Hz $↔$ 0.49 J/mol/rad$^2$.

Alternatively, the absolute magnitude of the magnetotropic coefficient $k$ of the sample can be estimated from \autoref{eq:freqshift}, which requires the bending stiffness of the fundamental mode of the lever, $K^{(0)}$. This was calculated in the methods section of \citet{Modic2018}, and is
\begin{align}\label{eq:effectiveK-akiyama}
K^{(0)}  = 180~\text{nJ}.
\end{align}
At 1 T, we observe a frequency shift of 0.48 Hz at 48 kHz. Using \autoref{eq:freqshift}, $k = 2 K^{(0)} (Δf/f_0) = $ 3.6 pJ/rad$^2$ at 1 T. The sample size is roughly $50×70×2$ μm$^3$, and with a density of $3.1$ g/cm$^3$, this yields 26 pmol (per unit cell containing 4 formula units). Thus, the specific (per mol) value of the magnetotropic coefficient is  $k$ = 3.6 pJ/rad$^2$ / 26 pmol = 0.14  J/mol/rad$^2$ at 1 T. In the linear regime, where $k=B^2(χ_1-χ_2)\cos2θ$, 0.14  J/mol/rad$^2$ can be compared to $k$ obtained from the known value of anisotropy in the magnetic susceptibility of 0.24 J/mol/T$^2$, leading to $k$ = 0.24  J/mol/rad$^2$ (because $B^2 = 1$ $\text{T}^2$ and $\cos2θ=1$). The main sources of uncertainty are the thickness of the sample and the estimated value of $K^{(0)}$. \autoref{eq:effectiveK-akiyama} describes a rectangular-shaped lever that is uniform along the length, whereas the levers  we use in the experiment have a more complex shape.

\section{Field-temperature scaling of the magnetotropic coefficient.}
\label{sec:scaling}

We use a mean-field model to illustrate the effects of exchange interactions on the $k/T$ versus $B/T$ scaling. The general Hamiltonian for a system of interacting spins is given by
\begin{align}\label{eq:hamiltonian}
H = -μ_B\B \cdot \g \cdot ∑_n (\bm{σ}^n/2) +  \frac14\,∑_{\av{n,m}} \bm{σ}^n \cdot \hat{J}^{nm} \cdot \bm{σ}^m\,,	
\end{align}
where $\bm{B}$ is the magnetic field, $\hat{g}$ is the $g-$factor tensor, $\hat{J}$ is the exchange matrix, $n$ and $m$ represent different lattice sites, and $\av{n,m}$ are nearest neighbors. We assume uniaxial $g-$factor anisotropy such that 
\begin{equation}
\hat{g} = \begin{pmatrix}    
g_a &0&0\\
0 & g_a & 0\\
0 & 0 & g_c
\end{pmatrix}.
\label{eq:g}
\end{equation}
\autoref{eq:g} is written in the basis of the crystal lattice (e.g. $a$, $b$, and $c^{\star}$, as in Figure 1B of the main text).

\begin{figure}[h!]
\includegraphics[width=.8\columnwidth]{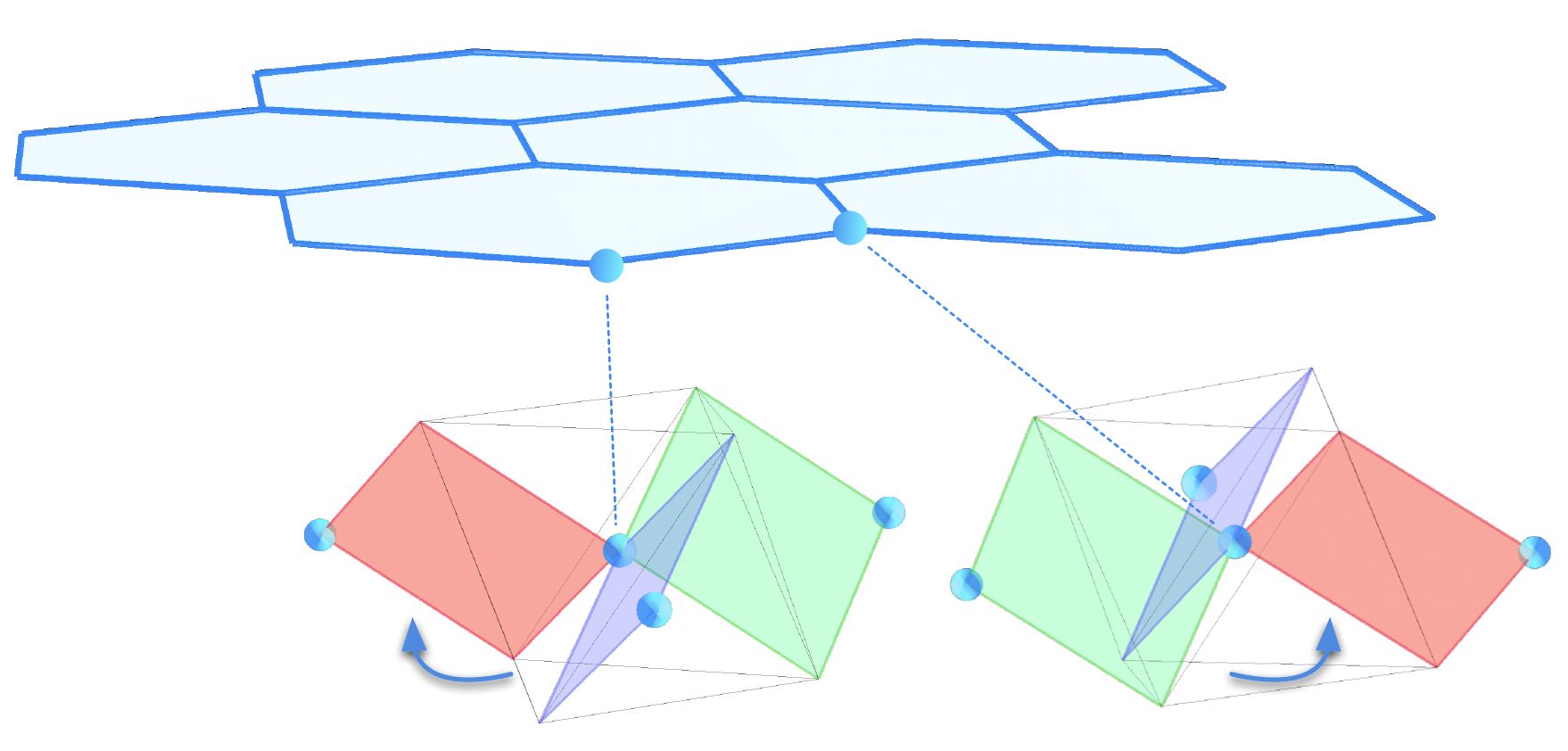}%
\singlespace
\caption{ \rucl\ has a planar honeycomb structure, where each Ru ion (blue spheres) is surrounded by Cl octahedra (Cl ions are located at the vertices of the octahedra, two of which are shown with thin black lines). All Cl octahedra in the structure are oriented identically. Two nearest-neighbor Ru ions share two Cl ions, which delineate the common edge shared between the neighboring octahedra. The exchange interaction between the two Ru ions is therefore mediated by two equivalent Cl ions simultaneously -- the interference between these two exchange paths is the microscopic basis for stronger spin-anisotropy of the exchange tensor $\hat{J}_{ij}$ on each Ru-Ru bond \cite{Jackeli2009}. Geometrically, these two exchange paths define a rectangular Ru-Cl-Ru-Cl ``exchange plaquette", highlighted here in blue, green, and red. The orientation of the plaquette defines three mutually orthogonal axes: two in the plane of the plaquette and one perpendicular to it. These directions are the principal components of the exchange tensor $J_{ij}$ on each Ru-Ru link, with the one perpendicular to the exchange plaquette describing the Kitaev component. The exchange plaquettes connecting the three nearest neighbors of each Ru ion are mutually orthogonal \cite{Jackeli2009}. Although the Cl octahedron is inversion-symmetric around the center Ru ion, this is not true when considering the three exchange plaquettes around each Ru ion. In particular, under a mirror reflection parallel to the honeycomb planes, the exchange environment on neighboring Ru ions  transform into each other. Overall, the exchange Hamiltonian describing the entire honeycomb plane of \rucl\ plane does not have reflection symmetry with respect to the plane parallel to itself unless $J_{ij}$ is isotropic.}%
\label{fig:water}%
\end{figure}

We consider spin-1/2's interacting with spin-anisotropic exchange on the bipartite lattice of RuCl$_3$.  Considering any pair of spins -- one on each sublattice -- the exchange between them across each Ru-Cl-Ru-Cl plaquette (\autoref{fig:water}) is 
\begin{align}\label{eq:plaquette}
\hat{J}_{\text{Ru-Cl-Ru-Cl}} = \sqmatr{ 
J_R	& 0 & 0\\
0 & J_C & 0\\
0 & 0 & J_K 
},
\end{align}
where it is important to note that this is \textit{not} the same basis as \autoref{eq:g}. Here the three orthogonal directions are along the Ru-Ru link, along the Cl-Cl direction, and the third is perpendicular to the Ru-Cl-Ru-Cl plaquette (see \autoref{fig:water}). In this basis, $J_K$ is the ``Kitaev'' exchange term. Components of the exchange tensor in the $a,b,c^{\star}$ basis tied to the honeycomb plane can be obtained by rotating $\hat{J}_{\text{Ru-Cl-Ru-Cl}}$ via $U_i^{\dagger} \cdot \hat{J}_{\text{Ru-Cl-Ru-Cl}}\cdot U_i$, where the matrices $U_{i=1,2,3}$ rotate by $\arccos(1/\sqrt3) \approx 55^\circ$ around each of the three Ru-Ru bonds. For each of the three nearest neighbors this yields
\begin{align}\label{eq:anisotropic-exchange}
\hat{J}_1 =& \sqmatr{
  \frac{2 J_K+J_C+J_R}{4} & \frac{2 J_K+J_C-3 J_R}{4 \sqrt{3}} & \frac{J_K-J_C}{\sqrt{6}} \\ 
 \frac{2 J_K+J_C-3 J_R}{4 \sqrt{3}} & \frac{2 J_K+J_C+9 J_R}{12} & \frac{J_K-J_C}{3 \sqrt{2}} \\  
 \frac{J_K-J_C}{\sqrt{6}} & \frac{J_K-J_C}{3 \sqrt{2}} & \frac{J_K+2 J_C}{3} 
 }\,,	\notag\\
\hat{J}_2 =& \sqmatr{
   J_R & 0 & 0 \\ 
 0 & \frac{2 J_K+J_C}{3} & \frac{\sqrt{2} (J_C-J_K)}{3} \\
 0 & \frac{\sqrt{2} (J_C-J_K)}{3}  & \frac{J_K+2 J_C}{3} 
 }\,,	\notag\\
\hat{J}_3 =& \sqmatr{
  \frac{2 J_K+J_C+J_R}{4} & -\frac{2 J_K+J_C-3 J_R}{4 \sqrt{3}} & \frac{J_C-J_K}{\sqrt{6}} \\
 -\frac{2 J_K+J_C-3 J_R}{4 \sqrt{3}} & \frac{2 J_K+J_C+9 J_R}{12} & \frac{J_K-J_C}{3 \sqrt{2}} \\
 \frac{J_C-J_K}{\sqrt{6}} & \frac{J_K-J_C}{3 \sqrt{2}} & \frac{J_K+2 J_C}{3}
 }\,.
\end{align}
This notation is equivalent to the standard notation used in e.g. \citet{Winter:2016} where the Heisenberg coupling is $J$, the Kitaev coupling is $K$, and the first-order off-diagonal coupling is $\Gamma$. The two notations are related with $J_R = J - \Gamma$, $J_K = J + K$, and $J_C = J + \Gamma$.

To solve \autoref{eq:hamiltonian} at a mean-field level, we consider the average exchange interaction between a single spin and its three neighbors, which all lie on the other sublattice. This average exchange is 
\begin{align}\label{eq:exchange-bipartite}
		\hat{J}_{\text{ave}} &= \frac{1}{3}\left(\hat{J}_1+\hat{J}_2+\hat{J}_3\right) \notag\\
		&= 		\frac{1}{3}\left(J_R+J_C+J_K\right)             		 \sqmatr{1&0&0\\0&1&0\\0&0&1}  
						+ \frac{J_R-J_C}{6}   		\sqmatr{1&0&0\\0&1&0\\0&0&-2} \\
						&=\frac{1}{3}	\sqmatr{J_R+J_C+J_K&0&0\\0&J_R+J_C+J_K&0\\0&0&2\left(J_C+J_K\right)}.
 \end{align}
In the notation of \citet{Winter:2016}, this is written as 
\begin{align}\label{eq:exchange2}
		\hat{J}_{\text{ave}} 		&=\frac{1}{3}	\sqmatr{\left(3J+K-\Gamma\right)&0&0\\0&\left(3J + K -\Gamma\right)&0\\0&0&\left(3J + K + 2\Gamma\right)}.
 \end{align}
It is important to note that the approximation of isotropic interactions between nearest-neighbors, which yields \autoref{eq:exchange2} and \autoref{eq:exchange-bipartite}, introduces a mirror symmetry in the honeycomb plane that takes $z\rightarrow -z$. The full microscopic Hamiltonian of RuCl$_3$, based on symmetry considerations and defined by exchange \autoref{eq:anisotropic-exchange}, lacks this symmetry.

We can now write \autoref{eq:hamiltonian} in a form where a single spin interacts with a mean-field magnetization through the average exchange interaction given in \autoref{eq:exchange2}:
\begin{align}\label{eq:hamiltonianMF}
H_{MF} = -μ_B\B \cdot \g \cdot ∑_n (\bm{σ}^n/2) + \frac{3}{N\mu_B}\hat{g}^{-1}\cdot\bm{M}\cdot\hat{J}_{\text{ave}}\cdot\sum_n(\bm{\sigma}^n/2),
\end{align}
where the magnetization $\bm{M} = N\mu_B\hat{g}\cdot\left<\bm{\sigma}/2\right>$. This equation can be solved by first calculating the free energy as
\begin{equation}
F = -k_B T \log\left(\mathrm{Tr} e^{-H_{MF}/k_B T}\right),
\label{eq:fen}
\end{equation}
and then the magnetization as 
\begin{equation}
\bm{M} = -\frac{\partial F}{\partial \bm{B}}.
\label{eq:mag}
\end{equation}
\autoref{eq:mag} can then be solved self-consistently by numerical evaluation for the magnetization at any given temperature, magnetic field strength, and magnetic field orientation. This magnetization is then inserted into \autoref{eq:fen}, and the magnetotropic coefficient is calculated as 
\begin{equation}
k = \frac{\partial^2 F}{\partial \theta^2}.
\label{eq:magnetotro}
\end{equation}

\begin{figure}[h!]
\includegraphics[width=.8\columnwidth]{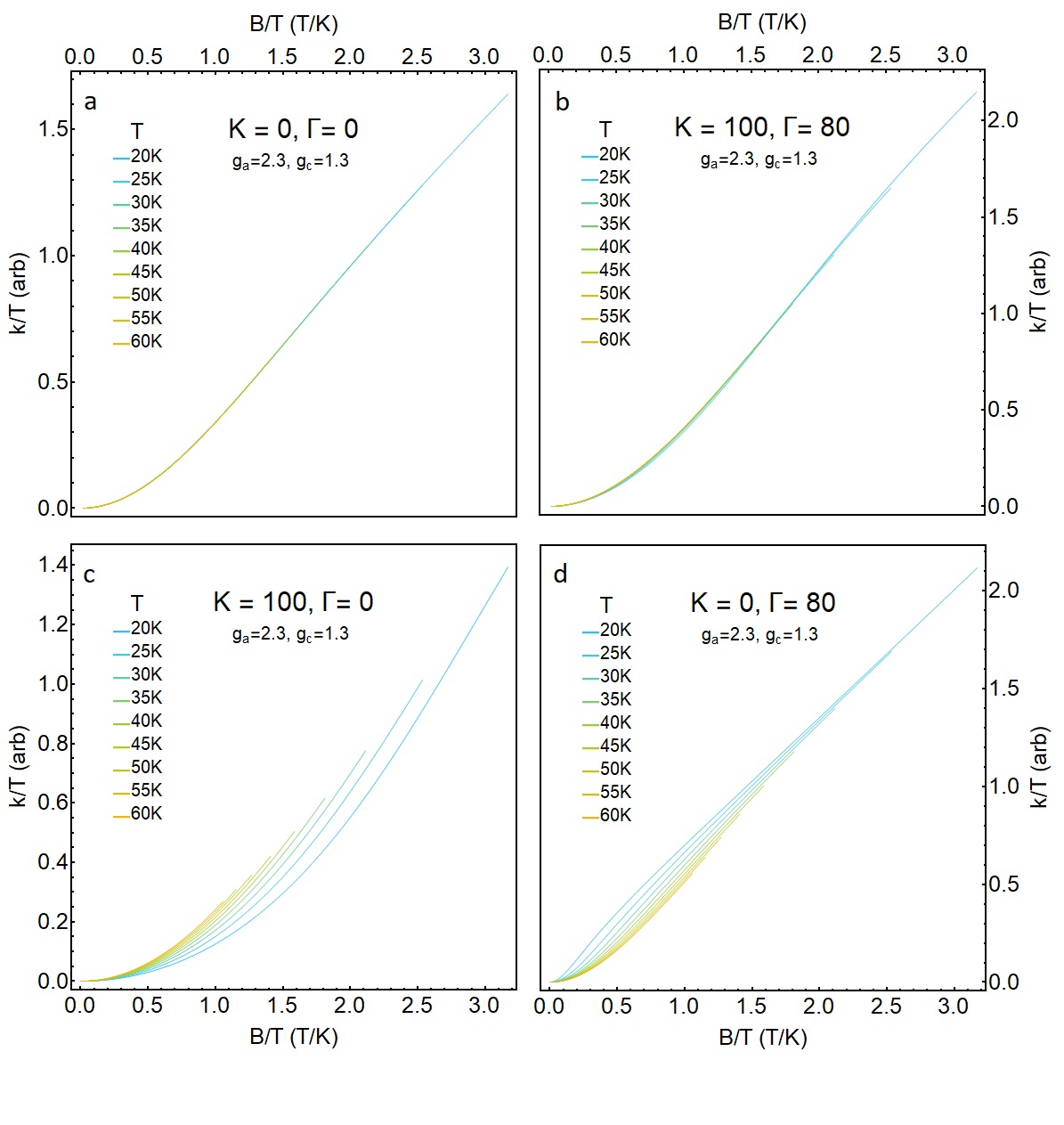}%
\singlespace
\caption{ The magnetotropic coefficient evaluated for a mean-field model with a $g-$factor and exchange as defined in \autoref{eq:hamiltonianMF}. $k/T$ is plotted as a function of $B/T$, with the magnetic field up to 63.5 T. Perfect scaling is achieved with no exchange interaction (a), but is destroyed by both on and off-diagonal exchange (c and d). The presence of both $K$ and $\Gamma$ restores scaling (b).   }%
\label{fig:sims}%
\end{figure}

\autoref{fig:sims} shows the magnetotropic coefficient calculated with this model for different values of the exchange interactions. The temperature range was chosen to coincide with that over which the data scales in Figure 3 of the main text. With no exchange (panel a), only field and temperature set the scale and thus $k/T$ scales when plotted versus $B/T$ at all fields and temperatures. When a finite $K$ or $\Gamma$ is introduced (panels c and d), however, the scaling is destroyed by the competing energy scale. When both $K$ and $\Gamma$ are present and of comparable magnitude (panel b), the data again scales.  

\section{The angle-dependent magnetotropic coefficient}
\label{sec:spike}

Figure 4 of the main text reveals a peculiar ``spike" in the angle dependence of the magnetotropic coefficient across the $c^⋆$ direction. In addition, each angle scan has an overall negative ``weight" (a finite negative area under $k(θ)$ over 180$^∘$). Both features strengthen with magnetic field and occur at temperatures both below and above the N\'eel temperature. In this section, we discuss these features in more detail, as they provide evidence for a thermodynamic singularity in the magnetic free energy of \rucl for fields near the $c^⋆$ direction. 

When the plane of rotation of the rotator stage is aligned with the plane of vibration of the lever, the torque $τ(θ)$ and the magnetotropic coefficient $k(θ)$ are related via $τ(θ)=∫^{θ}dθ'k(θ')$. Because $τ(θ)$ is a periodic function of $θ$ (with 180$^∘$ periodicity required by time-reversal symmetry, \autoref{sec:timerev}), the area under the angle scans of the magnetotropic coefficient over 180$^∘$ range, must be equal to zero $\av{k(θ)} = (1/π)∫_{θ}^{θ+π}dθ'k(θ')=0$. This is clearly not the case for the data in Figure 4, where the average $\av{k(θ)}$ over 180$^∘$ has a finite negative value.

The spike at $c^⋆$ hints at the non-zero $\av{k(θ)}$: the negative weight could be compensated by a larger (positive) area under the spike in order to recover 180$^∘$ periodicity. This implies that the measured area under the spike is not as large as expected and suggests a slight precession of the magnetic field about an extremely sharp singularity in the free energy at $c^⋆$. Both the spike and the negative weight independently point to this conclusion.

\begin{figure}[h!]
\includegraphics[width=.49\columnwidth]{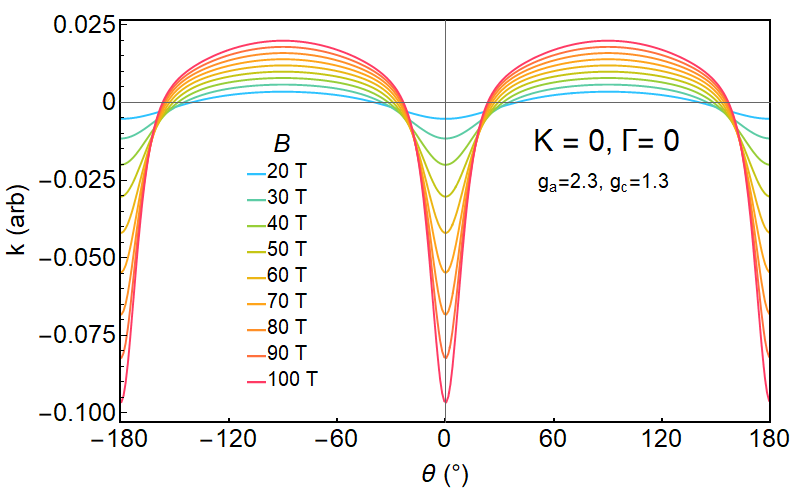}%
\includegraphics[width=.49\columnwidth]{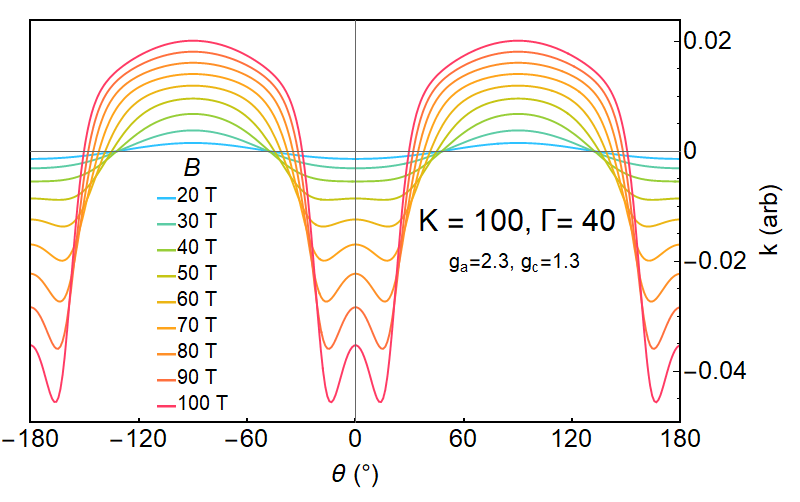}%
\singlespace
\caption{ The magnetotropic coefficient evaluated for a mean-field model, as a function of angle at $T = 10~$K. The left panel shows the angle dependence of $k$ for $g$-factor anisotropy alone. The large dip near $0^{\circ}$ is reminiscent of the data shown in Figure 4 of the main text. The spike, however, is only present once exchange interactions are present (right panel).  }%
\label{fig:sims2}%
\end{figure}

Using the mean field model from \autoref{sec:scaling} we plot the angle dependence of $k$ with and without exchange interactions (\autoref{fig:sims2}). $g-$factor anisotropy alone captures some features of the data but, in particular, does not show a ``spike'' near $c^{\star}$ for any values of $g_a$ and $g_c$. Introducing a sizable $K$ and $\Gamma$ produces a spike, although it is not possible to reproduce both the scale invariance as a function of field and the spike as a function of angle with the same set of parameters with this model. Note that the spike in the data of Figure 4 of the main text is also significantly narrower than what can be produced with both our mean field model or with a 24-site exact diagonalization \cite{Riedl:2019}.

\section{Time reversal check up to 64 T}
\label{sec:timerev}

\begin{figure}[h!]
\centerline{\includegraphics[width=1\columnwidth]{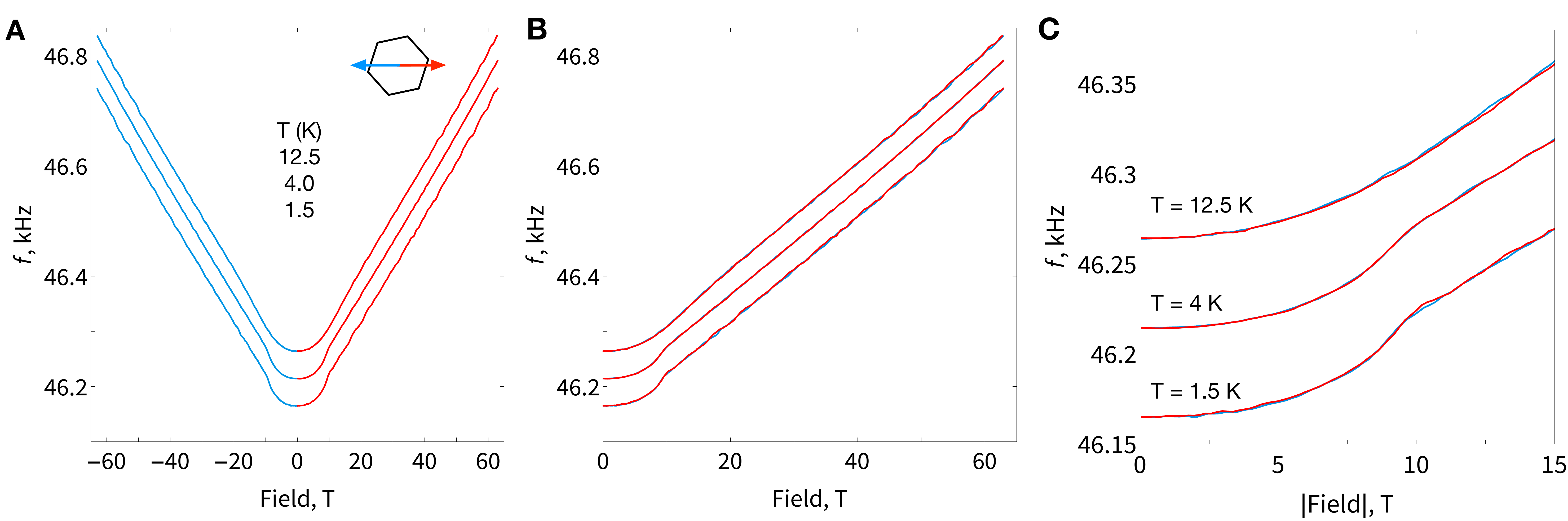}}
\singlespace
\caption{ Tests of time-reversal invariance of the magnetotropic coefficient for fields up to 64 T in the $ab$-plane. Data is taken for the same sample and orientation as Figure 3 of the main text and \autoref{fig:60T}. A) Full field scans at three different temperatures. The inset shows schematically the field orientation. The in-plane orientation of the magnetic field is unknown for the sample in Figure 3 of the main text. B) The field-reversed scan plotted versus the absolute value of magnetic field lays exactly over the positive field scans. C) Zoom-in of the low-field part of panel B.}
\label{fig:timereversal}
\end{figure}

The magnetic free energy $F(θ, φ)$ in a time-reversal-invariant system must be the same for opposite field orientations. This requires 180$^∘$ periodicity of the angle scans in the plane perpendicular to the honeycomb plane, and 60$^∘$ periodicity for angle scans in the plane parallel to the honeycomb planes. All angle scans at high fields in the plane perpendicular to the honeycomb plane show  180$^∘$ periodicity, indicating time-reversal-invariance in the high field state of RuCl$_3$. We also check this at higher magnetic fields by reversing the pulsed-magnetic-field direction. This is shown in \autoref{fig:timereversal}, which directly compares the frequency shift for two opposite field orientations in the $ab$-plane. Both at high and low fields, there is no experimental indication of a deviation from $B→ -B$ symmetry.

\section{Additional data}
\label{sec:adata}

Here we provide additional data that is used to construct the phase diagram of the AFM order as a function of field, temperature, and angle, shown in Figure 1c of the main text. 

\begin{figure}[h!]
\centering
\includegraphics[width=0.9\linewidth, trim=0cm 0cm 0cm 0cm, clip=true]{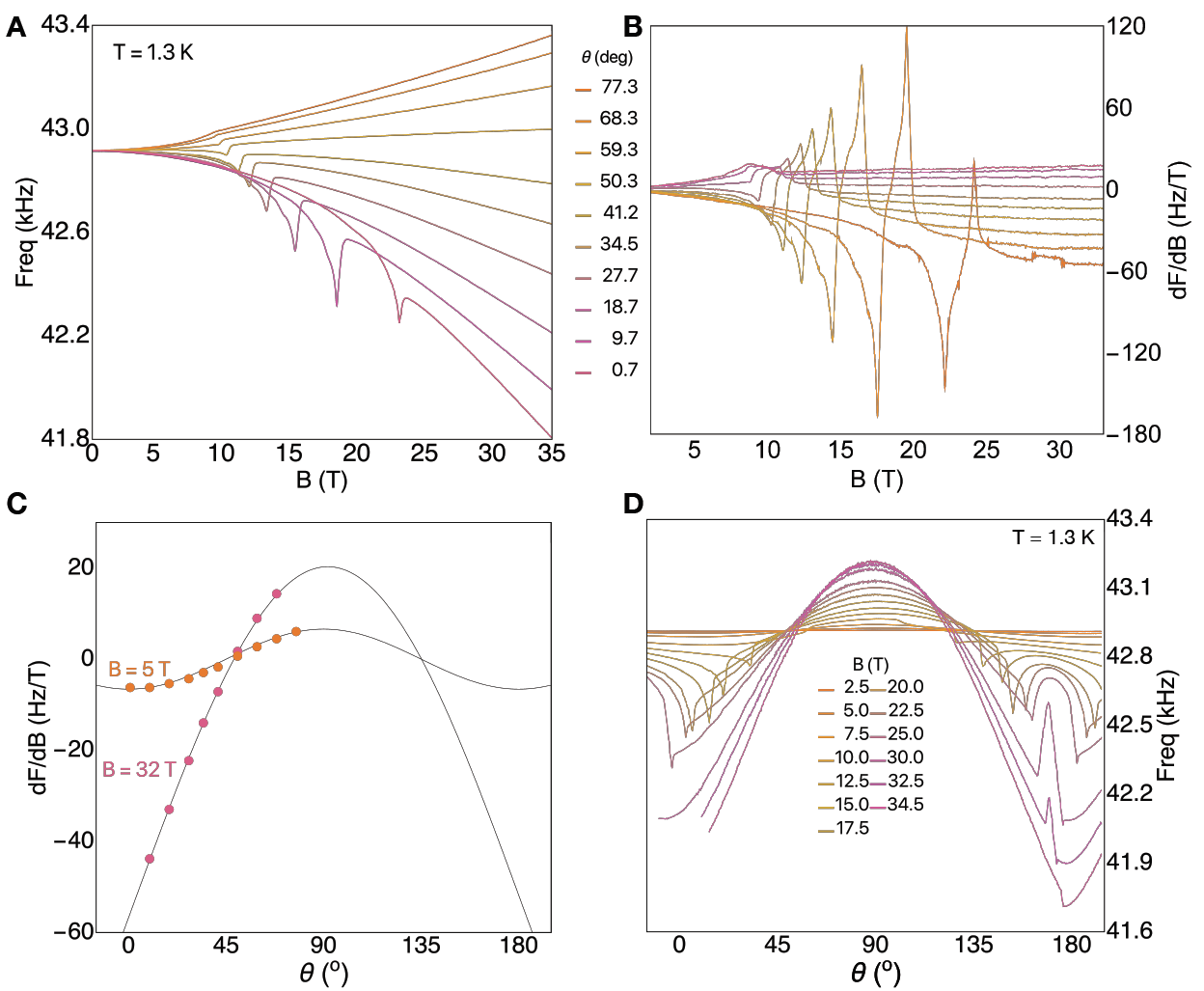}
\singlespace
\caption{ The magnetotropic coefficient of \rucl as a function of field and angle at 1.3 K. A) Frequency versus magnetic field for fields applied at various angles between the $c$-axis (0$^∘$) and the honeycomb planes (90$^∘$). B) The field-derivative of the data in A showing that the magnetotropic coefficient is constant above the AFM transition. C) The slope of the data in A determined for narrow field ranges around 5 and 32 T. The behavior agrees with D) the angle dependence of the magnetotropic coefficient under continuous rotation of the sample in magnetic fields ranging from 2.5 to 34.5 T.}
\label{fig:Figure_1K}
\end{figure}

\begin{figure}[h!]
\centering
\includegraphics[width=0.9\linewidth, trim=2cm 3cm 8cm 3cm, clip=true]{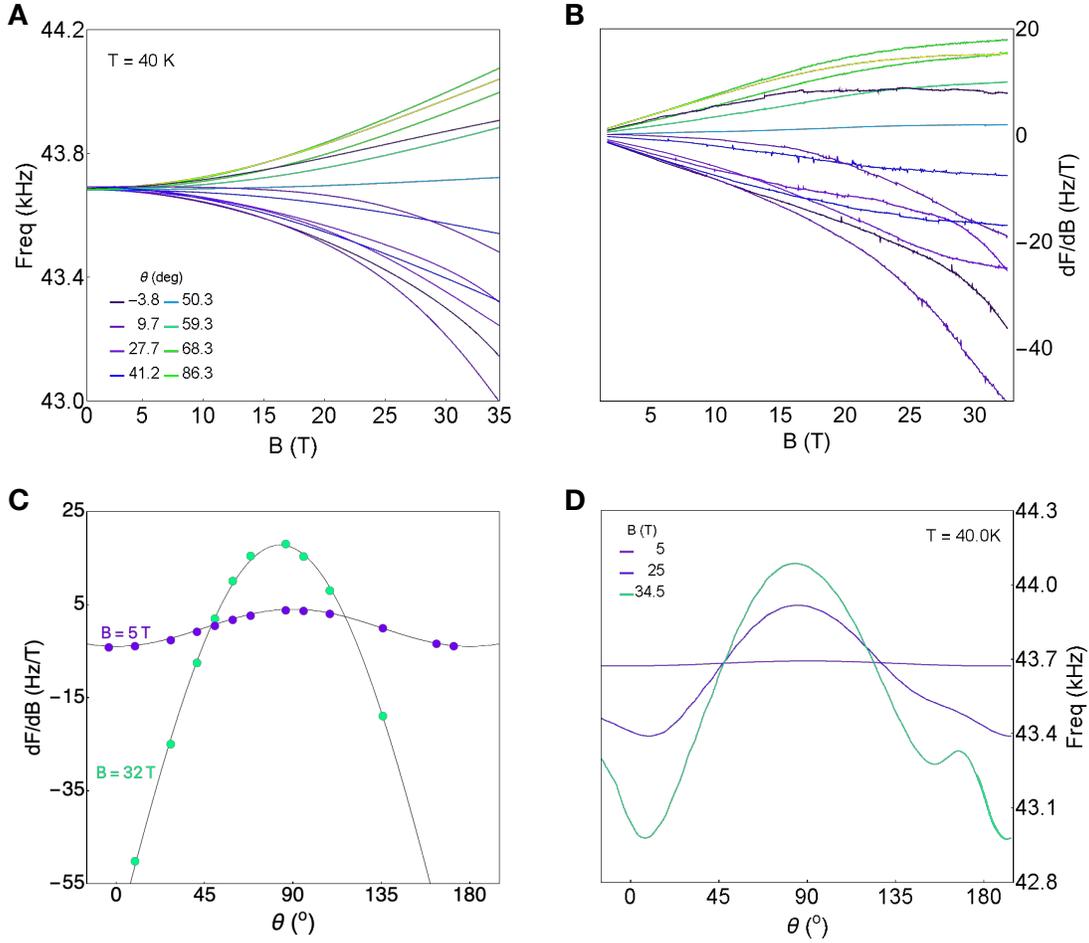}
\singlespace
\caption{ The magnetotropic coefficient of \rucl as a function of field and angle at 40 K. A) Frequency versus magnetic field for various angles ranging from near the $c$-axis (0\deg) and the honeycomb planes (90\deg). B) The field-derivative of the data in A shows an approach to saturation at high fields when magnetic field is applied near the honeycomb planes. For magnetic field applied near the $c$-axis, saturation is pushed to an inaccessible field range. C) The slope of the data in A determined for narrow field ranges around 5 and 32 T. The behavior is consistent with D) the angle dependence of the magnetotropic coefficient as the sample is continuously rotated in magnetic fields of 5, 25, and 34.5 T.}
\label{fig:Figure_40K}
\end{figure}

\begin{figure}[h!]
\centering
\includegraphics[width=.65\linewidth, trim=0cm 0cm 0cm 0cm, clip=true]{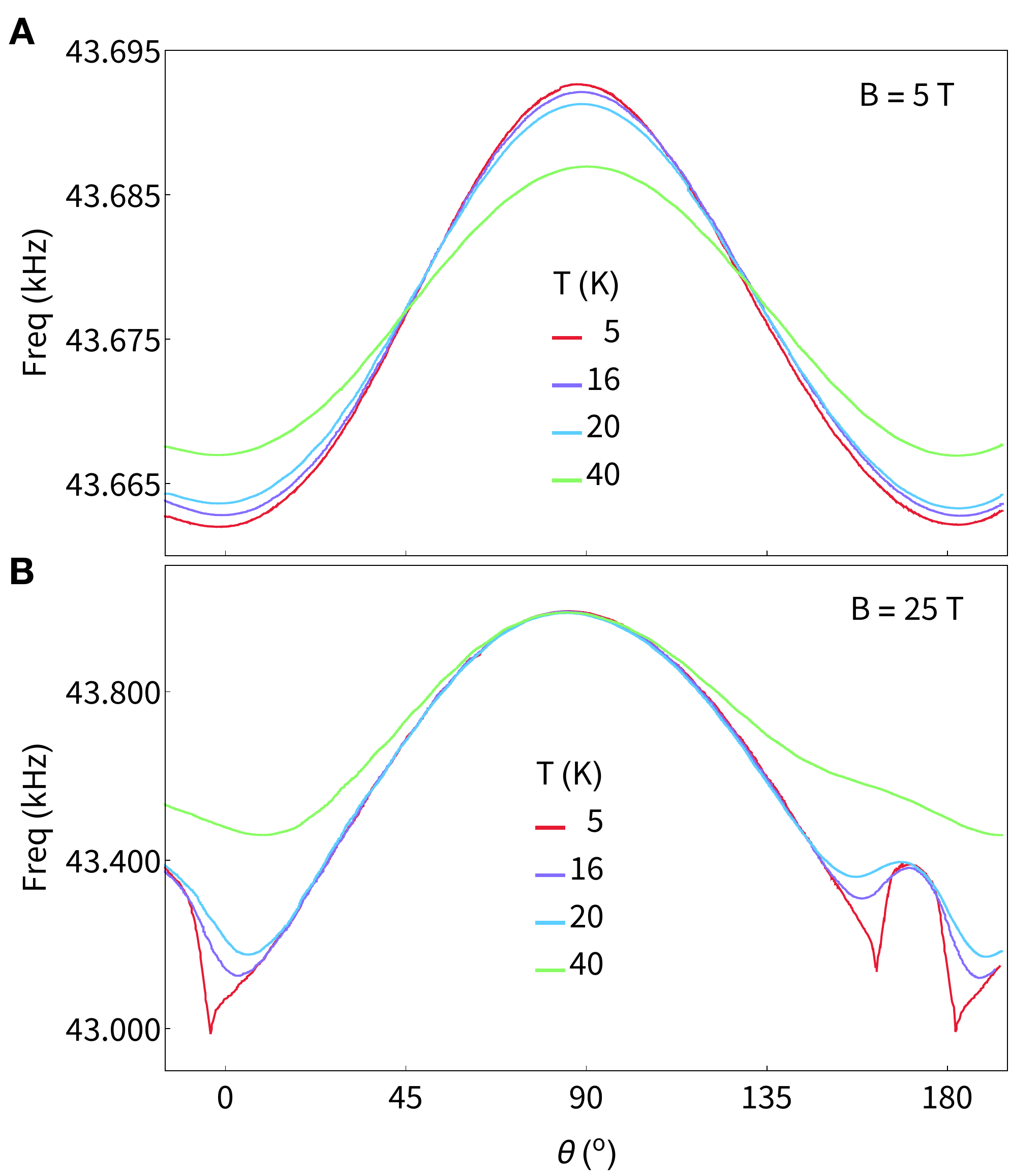}
\singlespace
\caption{ The magnetotropic coefficient of \rucl as the sample is rotated in fixed magnetic fields. A) Frequency shift (proportional to the magnetotropic coefficient) as a function of field orientation angle. $θ$ is defined from the $c$-axis (i.e., $θ$ = 90$\degm$ corresponds to magnetic field applied in the honeycomb planes). At 5 T, the expected $\cos2θ$ dependence is observed with a reduced amplitude (proportional to the anisotropy) upon increasing temperature. B) At 25 T and 5 K, entrance into (out of) the AFM phase occurs at $θ$ = 162$\degm$ ($θ$ = 182\deg). Signatures of the anisotropy that gives rise to the AFM phase are apparent at temperatures greater than 40 K.}
\label{fig:Figure_temps}
\end{figure}

\begin{figure}[h!]
\centering
\includegraphics[width=.65\linewidth, trim=0cm 0cm 0cm 0cm, clip=true]{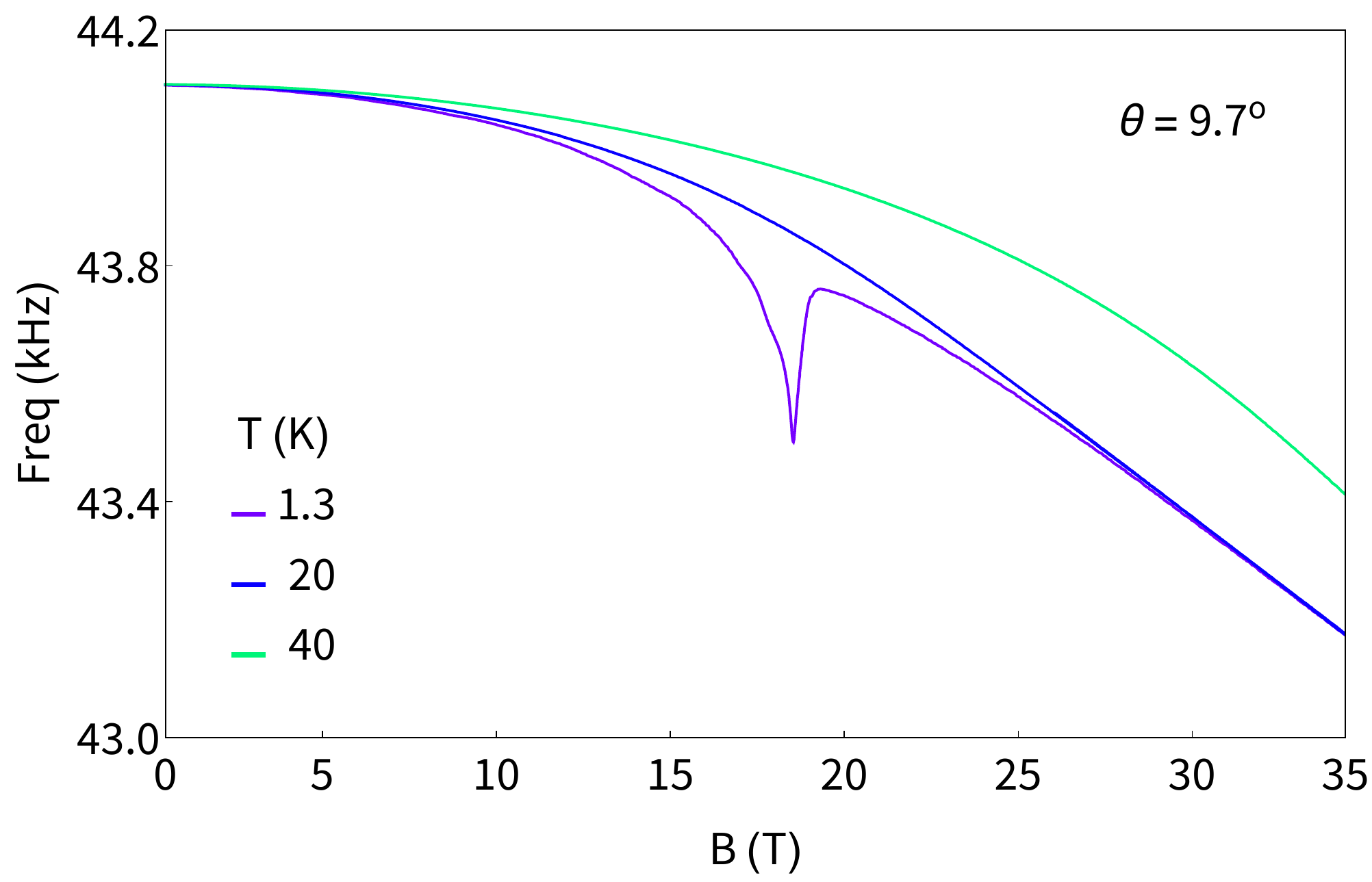}
\singlespace
\caption{ The magnetotropic coefficient of \rucl at $θ$ = 9.7$\degm$ for a few temperatures. At 1.3 K, the AFM phase boundary is traversed at roughly 18.5 T. At 20 K, the magnetotropic coefficient converges with the 1.3 K data above the AFM transition field. The high-field slope in the 40 K curve saturates to that of the lower temperature data, but with a larger field.}
\label{fig:Figure_angle}
\end{figure}


\begin{thebibliography}{37}%
\makeatletter
\providecommand \@ifxundefined [1]{%
 \@ifx{#1\undefined}
}%
\providecommand \@ifnum [1]{%
 \ifnum #1\expandafter \@firstoftwo
 \else \expandafter \@secondoftwo
 \fi
}%
\providecommand \@ifx [1]{%
 \ifx #1\expandafter \@firstoftwo
 \else \expandafter \@secondoftwo
 \fi
}%
\providecommand \natexlab [1]{#1}%
\providecommand \enquote  [1]{``#1''}%
\providecommand \bibnamefont  [1]{#1}%
\providecommand \bibfnamefont [1]{#1}%
\providecommand \citenamefont [1]{#1}%
\providecommand \href@noop [0]{\@secondoftwo}%
\providecommand \href [0]{\begingroup \@sanitize@url \@href}%
\providecommand \@href[1]{\@@startlink{#1}\@@href}%
\providecommand \@@href[1]{\endgroup#1\@@endlink}%
\providecommand \@sanitize@url [0]{\catcode `\\12\catcode `\$12\catcode
  `\&12\catcode `\#12\catcode `\^12\catcode `\_12\catcode `\%12\relax}%
\providecommand \@@startlink[1]{}%
\providecommand \@@endlink[0]{}%
\providecommand \url  [0]{\begingroup\@sanitize@url \@url }%
\providecommand \@url [1]{\endgroup\@href {#1}{\urlprefix }}%
\providecommand \urlprefix  [0]{URL }%
\providecommand \Eprint [0]{\href }%
\providecommand \doibase [0]{https://doi.org/}%
\providecommand \selectlanguage [0]{\@gobble}%
\providecommand \bibinfo  [0]{\@secondoftwo}%
\providecommand \bibfield  [0]{\@secondoftwo}%
\providecommand \translation [1]{[#1]}%
\providecommand \BibitemOpen [0]{}%
\providecommand \bibitemStop [0]{}%
\providecommand \bibitemNoStop [0]{.\EOS\space}%
\providecommand \EOS [0]{\spacefactor3000\relax}%
\providecommand \BibitemShut  [1]{\csname bibitem#1\endcsname}%
\let\auto@bib@innerbib\@empty
\bibitem [{\citenamefont {Kitaev}(2006)}]{Kitaev2006}%
  \BibitemOpen
  \bibfield  {author} {\bibinfo {author} {\bibfnamefont {A.}~\bibnamefont
  {Kitaev}},\ }\href {https://doi.org/10.1016/j.aop.2005.10.005} {\bibfield
  {journal} {\bibinfo  {journal} {{\it Annals of Physics }}\ }\textbf {\bibinfo
  {volume} {321}},\ \bibinfo {pages} {2} (\bibinfo {year} {2006})}\BibitemShut
  {NoStop}%
\bibitem [{\citenamefont {Kasahara}\ \emph
  {et~al.}(2018{\natexlab{a}})\citenamefont {Kasahara}, \citenamefont {Sugii},
  \citenamefont {Ohnishi}, \citenamefont {Shimozawa}, \citenamefont
  {Yamashita}, \citenamefont {Kurita}, \citenamefont {Tanaka}, \citenamefont
  {Nasu}, \citenamefont {Motome}, \citenamefont {Shibauchi} \emph
  {et~al.}}]{Kasahara2018}%
  \BibitemOpen
  \bibfield  {author} {\bibinfo {author} {\bibfnamefont {Y.}~\bibnamefont
  {Kasahara}}, \bibinfo {author} {\bibfnamefont {K.}~\bibnamefont {Sugii}},
  \bibinfo {author} {\bibfnamefont {T.}~\bibnamefont {Ohnishi}}, \bibinfo
  {author} {\bibfnamefont {M.}~\bibnamefont {Shimozawa}}, \bibinfo {author}
  {\bibfnamefont {M.}~\bibnamefont {Yamashita}}, \bibinfo {author}
  {\bibfnamefont {N.}~\bibnamefont {Kurita}}, \bibinfo {author} {\bibfnamefont
  {H.}~\bibnamefont {Tanaka}}, \bibinfo {author} {\bibfnamefont
  {J.}~\bibnamefont {Nasu}}, \bibinfo {author} {\bibfnamefont {Y.}~\bibnamefont
  {Motome}}, \bibinfo {author} {\bibfnamefont {T.}~\bibnamefont {Shibauchi}},
  \emph {et~al.},\ }\href {https://doi.org/10.1103/PhysRevLett.120.217205}
  {\bibfield  {journal} {\bibinfo  {journal} {{\it Physical Review Letters }}\
  }\textbf {\bibinfo {volume} {120}},\ \bibinfo {pages} {217205} (\bibinfo
  {year} {2018}{\natexlab{a}})}\BibitemShut {NoStop}%
\bibitem [{\citenamefont {Kasahara}\ \emph
  {et~al.}(2018{\natexlab{b}})\citenamefont {Kasahara}, \citenamefont
  {Ohnishi}, \citenamefont {Mizukami}, \citenamefont {Tanaka}, \citenamefont
  {Ma}, \citenamefont {Sugii}, \citenamefont {Kurita}, \citenamefont {Tanaka},
  \citenamefont {Nasu}, \citenamefont {Motome}, \citenamefont {Shibauchi},\
  and\ \citenamefont {Matsuda}}]{Kasahara2018b}%
  \BibitemOpen
  \bibfield  {author} {\bibinfo {author} {\bibfnamefont {Y.}~\bibnamefont
  {Kasahara}}, \bibinfo {author} {\bibfnamefont {T.}~\bibnamefont {Ohnishi}},
  \bibinfo {author} {\bibfnamefont {Y.}~\bibnamefont {Mizukami}}, \bibinfo
  {author} {\bibfnamefont {O.}~\bibnamefont {Tanaka}}, \bibinfo {author}
  {\bibfnamefont {S.}~\bibnamefont {Ma}}, \bibinfo {author} {\bibfnamefont
  {K.}~\bibnamefont {Sugii}}, \bibinfo {author} {\bibfnamefont
  {N.}~\bibnamefont {Kurita}}, \bibinfo {author} {\bibfnamefont
  {H.}~\bibnamefont {Tanaka}}, \bibinfo {author} {\bibfnamefont
  {J.}~\bibnamefont {Nasu}}, \bibinfo {author} {\bibfnamefont {Y.}~\bibnamefont
  {Motome}}, \bibinfo {author} {\bibfnamefont {T.}~\bibnamefont {Shibauchi}},\
  and\ \bibinfo {author} {\bibfnamefont {Y.}~\bibnamefont {Matsuda}},\ }\href
  {https://doi.org/10.1038/s41586-018-0274-0} {\bibfield  {journal} {\bibinfo
  {journal} {{\it Nature }}\ }\textbf {\bibinfo {volume} {559}},\ \bibinfo
  {pages} {227} (\bibinfo {year} {2018}{\natexlab{b}})}\BibitemShut {NoStop}%
\bibitem [{\citenamefont {Sears}\ \emph {et~al.}(2017)\citenamefont {Sears},
  \citenamefont {Zhao}, \citenamefont {Xu}, \citenamefont {Lynn},\ and\
  \citenamefont {Kim}}]{Sears2017}%
  \BibitemOpen
  \bibfield  {author} {\bibinfo {author} {\bibfnamefont {J.~A.}\ \bibnamefont
  {Sears}}, \bibinfo {author} {\bibfnamefont {Y.}~\bibnamefont {Zhao}},
  \bibinfo {author} {\bibfnamefont {Z.}~\bibnamefont {Xu}}, \bibinfo {author}
  {\bibfnamefont {J.~W.}\ \bibnamefont {Lynn}},\ and\ \bibinfo {author}
  {\bibfnamefont {Y.-J.}\ \bibnamefont {Kim}},\ }\href
  {https://doi.org/10.1103/PhysRevB.95.180411} {\bibfield  {journal} {\bibinfo
  {journal} {{\it Physical Review B }}\ }\textbf {\bibinfo {volume} {95}},\
  \bibinfo {pages} {180411} (\bibinfo {year} {2017})}\BibitemShut {NoStop}%
\bibitem [{\citenamefont {Wolter}\ \emph {et~al.}(2017)\citenamefont {Wolter},
  \citenamefont {Corredor}, \citenamefont {Janssen}, \citenamefont {Nenkov},
  \citenamefont {Sch{\"o}necker}, \citenamefont {Do}, \citenamefont {Choi},
  \citenamefont {Albrecht}, \citenamefont {Hunger}, \citenamefont {Doert},
  \citenamefont {Vojta},\ and\ \citenamefont {Buechner}}]{Wolter2017}%
  \BibitemOpen
  \bibfield  {author} {\bibinfo {author} {\bibfnamefont {A.~U.~B.}\
  \bibnamefont {Wolter}}, \bibinfo {author} {\bibfnamefont {L.~T.}\
  \bibnamefont {Corredor}}, \bibinfo {author} {\bibfnamefont {L.}~\bibnamefont
  {Janssen}}, \bibinfo {author} {\bibfnamefont {K.}~\bibnamefont {Nenkov}},
  \bibinfo {author} {\bibfnamefont {S.}~\bibnamefont {Sch{\"o}necker}},
  \bibinfo {author} {\bibfnamefont {S.-H.}\ \bibnamefont {Do}}, \bibinfo
  {author} {\bibfnamefont {K.-Y.}\ \bibnamefont {Choi}}, \bibinfo {author}
  {\bibfnamefont {R.}~\bibnamefont {Albrecht}}, \bibinfo {author}
  {\bibfnamefont {J.}~\bibnamefont {Hunger}}, \bibinfo {author} {\bibfnamefont
  {T.}~\bibnamefont {Doert}}, \bibinfo {author} {\bibfnamefont
  {M.}~\bibnamefont {Vojta}},\ and\ \bibinfo {author} {\bibfnamefont
  {B.}~\bibnamefont {Buechner}},\ }\href
  {https://doi.org/10.1103/PhysRevB.96.041405} {\bibfield  {journal} {\bibinfo
  {journal} {{\it Physical Review B }}\ }\textbf {\bibinfo {volume} {96}},\
  \bibinfo {pages} {041405} (\bibinfo {year} {2017})}\BibitemShut {NoStop}%
\bibitem [{\citenamefont {Sandilands}\ \emph {et~al.}(2015)\citenamefont
  {Sandilands}, \citenamefont {Tian}, \citenamefont {Plumb}, \citenamefont
  {Kim},\ and\ \citenamefont {Burch}}]{Sandilands2015}%
  \BibitemOpen
  \bibfield  {author} {\bibinfo {author} {\bibfnamefont {L.~J.}\ \bibnamefont
  {Sandilands}}, \bibinfo {author} {\bibfnamefont {Y.}~\bibnamefont {Tian}},
  \bibinfo {author} {\bibfnamefont {K.~W.}\ \bibnamefont {Plumb}}, \bibinfo
  {author} {\bibfnamefont {Y.-J.}\ \bibnamefont {Kim}},\ and\ \bibinfo {author}
  {\bibfnamefont {K.~S.}\ \bibnamefont {Burch}},\ }\href
  {https://doi.org/10.1103/PhysRevLett.114.147201} {\bibfield  {journal}
  {\bibinfo  {journal} {{\it Physical Review Letters }}\ }\textbf {\bibinfo
  {volume} {114}},\ \bibinfo {pages} {147201} (\bibinfo {year}
  {2015})}\BibitemShut {NoStop}%
\bibitem [{\citenamefont {Banerjee}\ \emph {et~al.}(2016)\citenamefont
  {Banerjee}, \citenamefont {Bridges}, \citenamefont {Yan}, \citenamefont
  {Aczel}, \citenamefont {Li}, \citenamefont {Stone}, \citenamefont {Granroth},
  \citenamefont {Lumsden}, \citenamefont {Yiu}, \citenamefont {Knolle},
  \citenamefont {Bhattacharjee}, \citenamefont {Kovrizhin}, \citenamefont
  {Moessner}, \citenamefont {Tennant}, \citenamefont {Mandrus},\ and\
  \citenamefont {Nagler}}]{Banerjee2016}%
  \BibitemOpen
  \bibfield  {author} {\bibinfo {author} {\bibfnamefont {A.}~\bibnamefont
  {Banerjee}}, \bibinfo {author} {\bibfnamefont {C.~A.}\ \bibnamefont
  {Bridges}}, \bibinfo {author} {\bibfnamefont {J.~Q.}\ \bibnamefont {Yan}},
  \bibinfo {author} {\bibfnamefont {A.~A.}\ \bibnamefont {Aczel}}, \bibinfo
  {author} {\bibfnamefont {L.}~\bibnamefont {Li}}, \bibinfo {author}
  {\bibfnamefont {M.~B.}\ \bibnamefont {Stone}}, \bibinfo {author}
  {\bibfnamefont {G.~E.}\ \bibnamefont {Granroth}}, \bibinfo {author}
  {\bibfnamefont {M.~D.}\ \bibnamefont {Lumsden}}, \bibinfo {author}
  {\bibfnamefont {Y.}~\bibnamefont {Yiu}}, \bibinfo {author} {\bibfnamefont
  {J.}~\bibnamefont {Knolle}}, \bibinfo {author} {\bibfnamefont
  {S.}~\bibnamefont {Bhattacharjee}}, \bibinfo {author} {\bibfnamefont {D.~L.}\
  \bibnamefont {Kovrizhin}}, \bibinfo {author} {\bibfnamefont {R.}~\bibnamefont
  {Moessner}}, \bibinfo {author} {\bibfnamefont {D.~A.}\ \bibnamefont
  {Tennant}}, \bibinfo {author} {\bibfnamefont {D.~G.}\ \bibnamefont
  {Mandrus}},\ and\ \bibinfo {author} {\bibfnamefont {S.~E.}\ \bibnamefont
  {Nagler}},\ }\href {https://doi.org/10.1038/nmat4604} {\bibfield  {journal}
  {\bibinfo  {journal} {{\it Nature Materials }}\ }\textbf {\bibinfo {volume}
  {15}},\ \bibinfo {pages} {733} (\bibinfo {year} {2016})}\BibitemShut
  {NoStop}%
\bibitem [{\citenamefont {Banerjee}\ \emph {et~al.}(2017)\citenamefont
  {Banerjee}, \citenamefont {Yan}, \citenamefont {Knolle}, \citenamefont
  {Bridges}, \citenamefont {Stone}, \citenamefont {Lumsden}, \citenamefont
  {Mandrus}, \citenamefont {Tennant}, \citenamefont {Moessner},\ and\
  \citenamefont {Nagler}}]{Banerjee2017}%
  \BibitemOpen
  \bibfield  {author} {\bibinfo {author} {\bibfnamefont {A.}~\bibnamefont
  {Banerjee}}, \bibinfo {author} {\bibfnamefont {J.}~\bibnamefont {Yan}},
  \bibinfo {author} {\bibfnamefont {J.}~\bibnamefont {Knolle}}, \bibinfo
  {author} {\bibfnamefont {C.~A.}\ \bibnamefont {Bridges}}, \bibinfo {author}
  {\bibfnamefont {M.~B.}\ \bibnamefont {Stone}}, \bibinfo {author}
  {\bibfnamefont {M.~D.}\ \bibnamefont {Lumsden}}, \bibinfo {author}
  {\bibfnamefont {D.~G.}\ \bibnamefont {Mandrus}}, \bibinfo {author}
  {\bibfnamefont {D.~A.}\ \bibnamefont {Tennant}}, \bibinfo {author}
  {\bibfnamefont {R.}~\bibnamefont {Moessner}},\ and\ \bibinfo {author}
  {\bibfnamefont {S.~E.}\ \bibnamefont {Nagler}},\ }\href
  {https://doi.org/10.1126/science.aah6015} {\bibfield  {journal} {\bibinfo
  {journal} {{\it Science }}\ }\textbf {\bibinfo {volume} {356}},\ \bibinfo
  {pages} {1055} (\bibinfo {year} {2017})}\BibitemShut {NoStop}%
\bibitem [{\citenamefont {Baek}\ \emph {et~al.}(2017)\citenamefont {Baek},
  \citenamefont {Do}, \citenamefont {Choi}, \citenamefont {Kwon}, \citenamefont
  {Wolter}, \citenamefont {Nishimoto}, \citenamefont {van~den Brink},\ and\
  \citenamefont {B{\"u}chner}}]{Baek2017}%
  \BibitemOpen
  \bibfield  {author} {\bibinfo {author} {\bibfnamefont {S.-H.}\ \bibnamefont
  {Baek}}, \bibinfo {author} {\bibfnamefont {S.-H.}\ \bibnamefont {Do}},
  \bibinfo {author} {\bibfnamefont {K.-Y.}\ \bibnamefont {Choi}}, \bibinfo
  {author} {\bibfnamefont {Y.}~\bibnamefont {Kwon}}, \bibinfo {author}
  {\bibfnamefont {A.}~\bibnamefont {Wolter}}, \bibinfo {author} {\bibfnamefont
  {S.}~\bibnamefont {Nishimoto}}, \bibinfo {author} {\bibfnamefont
  {J.}~\bibnamefont {van~den Brink}},\ and\ \bibinfo {author} {\bibfnamefont
  {B.}~\bibnamefont {B{\"u}chner}},\ }\href
  {https://doi.org/10.1103/PhysRevLett.119.037201} {\bibfield  {journal}
  {\bibinfo  {journal} {{\it Physical Review Letters }}\ }\textbf {\bibinfo
  {volume} {119}},\ \bibinfo {pages} {037201} (\bibinfo {year}
  {2017})}\BibitemShut {NoStop}%
\bibitem [{\citenamefont {Wang}\ \emph {et~al.}(2017)\citenamefont {Wang},
  \citenamefont {Reschke}, \citenamefont {H{\"u}vonen}, \citenamefont {Do},
  \citenamefont {Choi}, \citenamefont {Gensch}, \citenamefont {Nagel},
  \citenamefont {R{\~o}{\~o}m},\ and\ \citenamefont {Loidl}}]{Wang2017}%
  \BibitemOpen
  \bibfield  {author} {\bibinfo {author} {\bibfnamefont {Z.}~\bibnamefont
  {Wang}}, \bibinfo {author} {\bibfnamefont {S.}~\bibnamefont {Reschke}},
  \bibinfo {author} {\bibfnamefont {D.}~\bibnamefont {H{\"u}vonen}}, \bibinfo
  {author} {\bibfnamefont {S.-H.}\ \bibnamefont {Do}}, \bibinfo {author}
  {\bibfnamefont {K.-Y.}\ \bibnamefont {Choi}}, \bibinfo {author}
  {\bibfnamefont {M.}~\bibnamefont {Gensch}}, \bibinfo {author} {\bibfnamefont
  {U.}~\bibnamefont {Nagel}}, \bibinfo {author} {\bibfnamefont
  {T.}~\bibnamefont {R{\~o}{\~o}m}},\ and\ \bibinfo {author} {\bibfnamefont
  {A.}~\bibnamefont {Loidl}},\ }\href
  {https://doi.org/10.1103/PhysRevLett.119.227202} {\bibfield  {journal}
  {\bibinfo  {journal} {{\it Physical Review Letters }}\ }\textbf {\bibinfo
  {volume} {119}},\ \bibinfo {pages} {227202} (\bibinfo {year}
  {2017})}\BibitemShut {NoStop}%
\bibitem [{\citenamefont {Ponomaryov}\ \emph {et~al.}(2017)\citenamefont
  {Ponomaryov}, \citenamefont {Schulze}, \citenamefont {Wosnitza},
  \citenamefont {Lampen-Kelley}, \citenamefont {Banerjee}, \citenamefont {Yan},
  \citenamefont {Bridges}, \citenamefont {Mandrus}, \citenamefont {Nagler},
  \citenamefont {Kolezhuk},\ and\ \citenamefont {Zvyagin}}]{Ponomaryov2017}%
  \BibitemOpen
  \bibfield  {author} {\bibinfo {author} {\bibfnamefont {A.}~\bibnamefont
  {Ponomaryov}}, \bibinfo {author} {\bibfnamefont {E.}~\bibnamefont {Schulze}},
  \bibinfo {author} {\bibfnamefont {J.}~\bibnamefont {Wosnitza}}, \bibinfo
  {author} {\bibfnamefont {P.}~\bibnamefont {Lampen-Kelley}}, \bibinfo {author}
  {\bibfnamefont {A.}~\bibnamefont {Banerjee}}, \bibinfo {author}
  {\bibfnamefont {J.-Q.}\ \bibnamefont {Yan}}, \bibinfo {author} {\bibfnamefont
  {C.}~\bibnamefont {Bridges}}, \bibinfo {author} {\bibfnamefont
  {D.}~\bibnamefont {Mandrus}}, \bibinfo {author} {\bibfnamefont
  {S.}~\bibnamefont {Nagler}}, \bibinfo {author} {\bibfnamefont {A.~K.}\
  \bibnamefont {Kolezhuk}},\ and\ \bibinfo {author} {\bibfnamefont {S.~A.}\
  \bibnamefont {Zvyagin}},\ }\href {https://doi.org/10.1103/PhysRevB.96.241107}
  {\bibfield  {journal} {\bibinfo  {journal} {{\it Physical Review B }}\
  }\textbf {\bibinfo {volume} {96}},\ \bibinfo {pages} {241107} (\bibinfo
  {year} {2017})}\BibitemShut {NoStop}%
\bibitem [{\citenamefont {Jackeli}\ and\ \citenamefont
  {Khaliullin}(2009)}]{Jackeli2009}%
  \BibitemOpen
  \bibfield  {author} {\bibinfo {author} {\bibfnamefont {G.}~\bibnamefont
  {Jackeli}}\ and\ \bibinfo {author} {\bibfnamefont {G.}~\bibnamefont
  {Khaliullin}},\ }\href {https://doi.org/10.1103/PhysRevLett.102.017205}
  {\bibfield  {journal} {\bibinfo  {journal} {{\it Physical Review Letters }}\
  }\textbf {\bibinfo {volume} {102}},\ \bibinfo {pages} {017205} (\bibinfo
  {year} {2009})}\BibitemShut {NoStop}%
\bibitem [{\citenamefont {Savary}\ and\ \citenamefont
  {Balents}(2016)}]{Savary2016}%
  \BibitemOpen
  \bibfield  {author} {\bibinfo {author} {\bibfnamefont {L.}~\bibnamefont
  {Savary}}\ and\ \bibinfo {author} {\bibfnamefont {L.}~\bibnamefont
  {Balents}},\ }\href {https://doi.org/10.1088/0034-4885/80/1/016502}
  {\bibfield  {journal} {\bibinfo  {journal} {{\it Reports on Progress in
  Physics }}\ }\textbf {\bibinfo {volume} {80}},\ \bibinfo {pages} {016502}
  (\bibinfo {year} {2016})}\BibitemShut {NoStop}%
\bibitem [{\citenamefont {Modic}\ \emph {et~al.}(2014)\citenamefont {Modic},
  \citenamefont {Smidt}, \citenamefont {Kimchi}, \citenamefont {Breznay},
  \citenamefont {Biffin}, \citenamefont {Choi}, \citenamefont {Johnson},
  \citenamefont {Coldea}, \citenamefont {Watkins-Curry}, \citenamefont
  {McCandless}, \citenamefont {Chan}, \citenamefont {Gandara}, \citenamefont
  {Islam}, \citenamefont {Vishwanath}, \citenamefont {Shekhter}, \citenamefont
  {McDonald},\ and\ \citenamefont {Analytis}}]{Modic2014}%
  \BibitemOpen
  \bibfield  {author} {\bibinfo {author} {\bibfnamefont {K.~A.}\ \bibnamefont
  {Modic}}, \bibinfo {author} {\bibfnamefont {T.~E.}\ \bibnamefont {Smidt}},
  \bibinfo {author} {\bibfnamefont {I.}~\bibnamefont {Kimchi}}, \bibinfo
  {author} {\bibfnamefont {N.~P.}\ \bibnamefont {Breznay}}, \bibinfo {author}
  {\bibfnamefont {A.}~\bibnamefont {Biffin}}, \bibinfo {author} {\bibfnamefont
  {S.}~\bibnamefont {Choi}}, \bibinfo {author} {\bibfnamefont {R.~D.}\
  \bibnamefont {Johnson}}, \bibinfo {author} {\bibfnamefont {R.}~\bibnamefont
  {Coldea}}, \bibinfo {author} {\bibfnamefont {P.}~\bibnamefont
  {Watkins-Curry}}, \bibinfo {author} {\bibfnamefont {G.~T.}\ \bibnamefont
  {McCandless}}, \bibinfo {author} {\bibfnamefont {J.~Y.}\ \bibnamefont
  {Chan}}, \bibinfo {author} {\bibfnamefont {F.}~\bibnamefont {Gandara}},
  \bibinfo {author} {\bibfnamefont {Z.}~\bibnamefont {Islam}}, \bibinfo
  {author} {\bibfnamefont {A.}~\bibnamefont {Vishwanath}}, \bibinfo {author}
  {\bibfnamefont {A.}~\bibnamefont {Shekhter}}, \bibinfo {author}
  {\bibfnamefont {R.~D.}\ \bibnamefont {McDonald}},\ and\ \bibinfo {author}
  {\bibfnamefont {J.~G.}\ \bibnamefont {Analytis}},\ }\href
  {https://doi.org/10.1038/ncomms5203} {\bibfield  {journal} {\bibinfo
  {journal} {{\it Nature Communications }}\ }\textbf {\bibinfo {volume} {5}},\
  \bibinfo {pages} {4203} (\bibinfo {year} {2014})}\BibitemShut {NoStop}%
\bibitem [{\citenamefont {Chun}\ \emph {et~al.}(2015)\citenamefont {Chun},
  \citenamefont {Kim}, \citenamefont {Kim}, \citenamefont {Zheng},
  \citenamefont {Stoumpos}, \citenamefont {Malliakas}, \citenamefont
  {Mitchell}, \citenamefont {Mehlawat}, \citenamefont {Singh}, \citenamefont
  {Choi}, \citenamefont {Gog}, \citenamefont {Al-Zein}, \citenamefont
  {Moretti~Sala}, \citenamefont {Krisch}, \citenamefont {Chaloupka},
  \citenamefont {Jackeli}, \citenamefont {Khaliullin},\ and\ \citenamefont
  {Kim}}]{Chun2015}%
  \BibitemOpen
  \bibfield  {author} {\bibinfo {author} {\bibfnamefont {S.~H.}\ \bibnamefont
  {Chun}}, \bibinfo {author} {\bibfnamefont {J.-W.}\ \bibnamefont {Kim}},
  \bibinfo {author} {\bibfnamefont {J.}~\bibnamefont {Kim}}, \bibinfo {author}
  {\bibfnamefont {H.}~\bibnamefont {Zheng}}, \bibinfo {author} {\bibfnamefont
  {C.~C.}\ \bibnamefont {Stoumpos}}, \bibinfo {author} {\bibfnamefont
  {C.}~\bibnamefont {Malliakas}}, \bibinfo {author} {\bibfnamefont
  {J.}~\bibnamefont {Mitchell}}, \bibinfo {author} {\bibfnamefont
  {K.}~\bibnamefont {Mehlawat}}, \bibinfo {author} {\bibfnamefont
  {Y.}~\bibnamefont {Singh}}, \bibinfo {author} {\bibfnamefont
  {Y.}~\bibnamefont {Choi}}, \bibinfo {author} {\bibfnamefont {T.}~\bibnamefont
  {Gog}}, \bibinfo {author} {\bibfnamefont {A.}~\bibnamefont {Al-Zein}},
  \bibinfo {author} {\bibfnamefont {M.}~\bibnamefont {Moretti~Sala}}, \bibinfo
  {author} {\bibfnamefont {M.}~\bibnamefont {Krisch}}, \bibinfo {author}
  {\bibfnamefont {J.}~\bibnamefont {Chaloupka}}, \bibinfo {author}
  {\bibfnamefont {G.}~\bibnamefont {Jackeli}}, \bibinfo {author} {\bibfnamefont
  {G.}~\bibnamefont {Khaliullin}},\ and\ \bibinfo {author} {\bibfnamefont
  {B.}~\bibnamefont {Kim}},\ }\href@noop {} {\bibfield  {journal} {\bibinfo
  {journal} {Nature Physics}\ }\textbf {\bibinfo {volume} {11}},\ \bibinfo
  {pages} {462} (\bibinfo {year} {2015})}\BibitemShut {NoStop}%
\bibitem [{\citenamefont {Majumder}\ \emph {et~al.}(2015)\citenamefont
  {Majumder}, \citenamefont {Schmidt}, \citenamefont {Rosner}, \citenamefont
  {Tsirlin}, \citenamefont {Yasuoka},\ and\ \citenamefont
  {Baenitz}}]{Majumder2015}%
  \BibitemOpen
  \bibfield  {author} {\bibinfo {author} {\bibfnamefont {M.}~\bibnamefont
  {Majumder}}, \bibinfo {author} {\bibfnamefont {M.}~\bibnamefont {Schmidt}},
  \bibinfo {author} {\bibfnamefont {H.}~\bibnamefont {Rosner}}, \bibinfo
  {author} {\bibfnamefont {A.}~\bibnamefont {Tsirlin}}, \bibinfo {author}
  {\bibfnamefont {H.}~\bibnamefont {Yasuoka}},\ and\ \bibinfo {author}
  {\bibfnamefont {M.}~\bibnamefont {Baenitz}},\ }\href
  {https://doi.org/10.1103/PhysRevB.91.180401} {\bibfield  {journal} {\bibinfo
  {journal} {{\it Physical Review B }}\ }\textbf {\bibinfo {volume} {91}},\
  \bibinfo {pages} {180401} (\bibinfo {year} {2015})}\BibitemShut {NoStop}%
\bibitem [{\citenamefont {Kimchi}\ \emph {et~al.}(2014)\citenamefont {Kimchi},
  \citenamefont {Analytis},\ and\ \citenamefont {Vishwanath}}]{Kimchi2014}%
  \BibitemOpen
  \bibfield  {author} {\bibinfo {author} {\bibfnamefont {I.}~\bibnamefont
  {Kimchi}}, \bibinfo {author} {\bibfnamefont {J.~G.}\ \bibnamefont
  {Analytis}},\ and\ \bibinfo {author} {\bibfnamefont {A.}~\bibnamefont
  {Vishwanath}},\ }\href {https://doi.org/10.1103/PhysRevB.90.205126}
  {\bibfield  {journal} {\bibinfo  {journal} {{\it Physical Review B }}\
  }\textbf {\bibinfo {volume} {90}},\ \bibinfo {pages} {205126} (\bibinfo
  {year} {2014})}\BibitemShut {NoStop}%
\bibitem [{\citenamefont {Chaloupka}\ \emph {et~al.}(2013)\citenamefont
  {Chaloupka}, \citenamefont {Jackeli},\ and\ \citenamefont
  {Khaliullin}}]{Chaloupka2013}%
  \BibitemOpen
  \bibfield  {author} {\bibinfo {author} {\bibfnamefont {J.}~\bibnamefont
  {Chaloupka}}, \bibinfo {author} {\bibfnamefont {G.}~\bibnamefont {Jackeli}},\
  and\ \bibinfo {author} {\bibfnamefont {G.}~\bibnamefont {Khaliullin}},\
  }\href {https://doi.org/10.1103/PhysRevLett.110.097204} {\bibfield  {journal}
  {\bibinfo  {journal} {{\it Physical Review Letters }}\ }\textbf {\bibinfo
  {volume} {110}},\ \bibinfo {pages} {097204} (\bibinfo {year}
  {2013})}\BibitemShut {NoStop}%
\bibitem [{\citenamefont {Das}\ \emph {et~al.}(2019)\citenamefont {Das},
  \citenamefont {Kundu}, \citenamefont {Zhu}, \citenamefont {Mun},
  \citenamefont {McDonald}, \citenamefont {Li}, \citenamefont {Balicas},
  \citenamefont {McCollam}, \citenamefont {Cao}, \citenamefont {Rau} \emph
  {et~al.}}]{Das:2019}%
  \BibitemOpen
  \bibfield  {author} {\bibinfo {author} {\bibfnamefont {S.~D.}\ \bibnamefont
  {Das}}, \bibinfo {author} {\bibfnamefont {S.}~\bibnamefont {Kundu}}, \bibinfo
  {author} {\bibfnamefont {Z.}~\bibnamefont {Zhu}}, \bibinfo {author}
  {\bibfnamefont {E.}~\bibnamefont {Mun}}, \bibinfo {author} {\bibfnamefont
  {R.~D.}\ \bibnamefont {McDonald}}, \bibinfo {author} {\bibfnamefont
  {G.}~\bibnamefont {Li}}, \bibinfo {author} {\bibfnamefont {L.}~\bibnamefont
  {Balicas}}, \bibinfo {author} {\bibfnamefont {A.}~\bibnamefont {McCollam}},
  \bibinfo {author} {\bibfnamefont {G.}~\bibnamefont {Cao}}, \bibinfo {author}
  {\bibfnamefont {J.~G.}\ \bibnamefont {Rau}}, \emph {et~al.},\ }\href@noop {}
  {\bibfield  {journal} {\bibinfo  {journal} {Physical Review B}\ }\textbf
  {\bibinfo {volume} {99}},\ \bibinfo {pages} {081101} (\bibinfo {year}
  {2019})}\BibitemShut {NoStop}%
\bibitem [{\citenamefont {Johnson}\ \emph {et~al.}(2015)\citenamefont
  {Johnson}, \citenamefont {Williams}, \citenamefont {Haghighirad},
  \citenamefont {Singleton}, \citenamefont {Zapf}, \citenamefont {Manuel},
  \citenamefont {Mazin}, \citenamefont {Li}, \citenamefont {Jeschke},
  \citenamefont {Valenti},\ and\ \citenamefont {Coldea}}]{Johnson2015}%
  \BibitemOpen
  \bibfield  {author} {\bibinfo {author} {\bibfnamefont {R.~D.}\ \bibnamefont
  {Johnson}}, \bibinfo {author} {\bibfnamefont {S.~C.}\ \bibnamefont
  {Williams}}, \bibinfo {author} {\bibfnamefont {A.~A.}\ \bibnamefont
  {Haghighirad}}, \bibinfo {author} {\bibfnamefont {J.}~\bibnamefont
  {Singleton}}, \bibinfo {author} {\bibfnamefont {V.}~\bibnamefont {Zapf}},
  \bibinfo {author} {\bibfnamefont {P.}~\bibnamefont {Manuel}}, \bibinfo
  {author} {\bibfnamefont {I.~I.}\ \bibnamefont {Mazin}}, \bibinfo {author}
  {\bibfnamefont {Y.}~\bibnamefont {Li}}, \bibinfo {author} {\bibfnamefont
  {H.~O.}\ \bibnamefont {Jeschke}}, \bibinfo {author} {\bibfnamefont
  {R.}~\bibnamefont {Valenti}},\ and\ \bibinfo {author} {\bibfnamefont
  {R.}~\bibnamefont {Coldea}},\ }\href
  {https://doi.org/10.1103/PhysRevB.92.235119} {\bibfield  {journal} {\bibinfo
  {journal} {{\it Physical Review B }}\ }\textbf {\bibinfo {volume} {92}},\
  \bibinfo {pages} {235119} (\bibinfo {year} {2015})}\BibitemShut {NoStop}%
\bibitem [{\citenamefont {Kubota}\ \emph {et~al.}(2015)\citenamefont {Kubota},
  \citenamefont {Tanaka}, \citenamefont {Ono}, \citenamefont {Narumi},\ and\
  \citenamefont {Kindo}}]{Kubota2015}%
  \BibitemOpen
  \bibfield  {author} {\bibinfo {author} {\bibfnamefont {Y.}~\bibnamefont
  {Kubota}}, \bibinfo {author} {\bibfnamefont {H.}~\bibnamefont {Tanaka}},
  \bibinfo {author} {\bibfnamefont {T.}~\bibnamefont {Ono}}, \bibinfo {author}
  {\bibfnamefont {Y.}~\bibnamefont {Narumi}},\ and\ \bibinfo {author}
  {\bibfnamefont {K.}~\bibnamefont {Kindo}},\ }\href
  {https://doi.org/10.1103/PhysRevB.91.094422} {\bibfield  {journal} {\bibinfo
  {journal} {{\it Physical Review B}}\ }\textbf {\bibinfo {volume} {91}},\
  \bibinfo {pages} {094422} (\bibinfo {year} {2015})}\BibitemShut {NoStop}%
\bibitem [{\citenamefont {Cao}\ \emph {et~al.}(2016)\citenamefont {Cao},
  \citenamefont {Banerjee}, \citenamefont {Yan}, \citenamefont {Bridges},
  \citenamefont {Lumsden}, \citenamefont {Mandrus}, \citenamefont {Tennant},
  \citenamefont {Chakoumakos},\ and\ \citenamefont {Nagler}}]{Cao2016}%
  \BibitemOpen
  \bibfield  {author} {\bibinfo {author} {\bibfnamefont {H.~B.}\ \bibnamefont
  {Cao}}, \bibinfo {author} {\bibfnamefont {A.}~\bibnamefont {Banerjee}},
  \bibinfo {author} {\bibfnamefont {J.-Q.}\ \bibnamefont {Yan}}, \bibinfo
  {author} {\bibfnamefont {C.~A.}\ \bibnamefont {Bridges}}, \bibinfo {author}
  {\bibfnamefont {M.~D.}\ \bibnamefont {Lumsden}}, \bibinfo {author}
  {\bibfnamefont {D.~G.}\ \bibnamefont {Mandrus}}, \bibinfo {author}
  {\bibfnamefont {D.~A.}\ \bibnamefont {Tennant}}, \bibinfo {author}
  {\bibfnamefont {B.~C.}\ \bibnamefont {Chakoumakos}},\ and\ \bibinfo {author}
  {\bibfnamefont {S.~E.}\ \bibnamefont {Nagler}},\ }\href
  {https://doi.org/10.1103/PhysRevB.93.134423} {\bibfield  {journal} {\bibinfo
  {journal} {{\it Physical Review B }}\ }\textbf {\bibinfo {volume} {93}},\
  \bibinfo {pages} {134423} (\bibinfo {year} {2016})}\BibitemShut {NoStop}%
\bibitem [{\citenamefont {Modic}\ \emph {et~al.}(2017)\citenamefont {Modic},
  \citenamefont {Ramshaw}, \citenamefont {Betts}, \citenamefont {Breznay},
  \citenamefont {Analytis}, \citenamefont {McDonald},\ and\ \citenamefont
  {Shekhter}}]{Modic2017}%
  \BibitemOpen
  \bibfield  {author} {\bibinfo {author} {\bibfnamefont {K.~A.}\ \bibnamefont
  {Modic}}, \bibinfo {author} {\bibfnamefont {B.~J.}\ \bibnamefont {Ramshaw}},
  \bibinfo {author} {\bibfnamefont {J.~B.}\ \bibnamefont {Betts}}, \bibinfo
  {author} {\bibfnamefont {N.~P.}\ \bibnamefont {Breznay}}, \bibinfo {author}
  {\bibfnamefont {J.~G.}\ \bibnamefont {Analytis}}, \bibinfo {author}
  {\bibfnamefont {R.~D.}\ \bibnamefont {McDonald}},\ and\ \bibinfo {author}
  {\bibfnamefont {A.}~\bibnamefont {Shekhter}},\ }\href
  {https://doi.org/10.1038/s41467-017-00264-6} {\bibfield  {journal} {\bibinfo
  {journal} {{\it Nature Communications }}\ }\textbf {\bibinfo {volume} {8}},\
  \bibinfo {pages} {180} (\bibinfo {year} {2017})}\BibitemShut {NoStop}%
\bibitem [{\citenamefont {Leahy}\ \emph {et~al.}(2017)\citenamefont {Leahy},
  \citenamefont {Pocs}, \citenamefont {Siegfried}, \citenamefont {Graf},
  \citenamefont {Do}, \citenamefont {Choi}, \citenamefont {Normand},\ and\
  \citenamefont {Lee}}]{Leahy2017}%
  \BibitemOpen
  \bibfield  {author} {\bibinfo {author} {\bibfnamefont {I.~A.}\ \bibnamefont
  {Leahy}}, \bibinfo {author} {\bibfnamefont {C.~A.}\ \bibnamefont {Pocs}},
  \bibinfo {author} {\bibfnamefont {P.~E.}\ \bibnamefont {Siegfried}}, \bibinfo
  {author} {\bibfnamefont {D.}~\bibnamefont {Graf}}, \bibinfo {author}
  {\bibfnamefont {S.-H.}\ \bibnamefont {Do}}, \bibinfo {author} {\bibfnamefont
  {K.-Y.}\ \bibnamefont {Choi}}, \bibinfo {author} {\bibfnamefont
  {B.}~\bibnamefont {Normand}},\ and\ \bibinfo {author} {\bibfnamefont
  {M.}~\bibnamefont {Lee}},\ }\href@noop {} {\bibfield  {journal} {\bibinfo
  {journal} {Physical review letters}\ }\textbf {\bibinfo {volume} {118}},\
  \bibinfo {pages} {187203} (\bibinfo {year} {2017})}\BibitemShut {NoStop}%
\bibitem [{\citenamefont {Yoshitake}\ \emph {et~al.}(2020)\citenamefont
  {Yoshitake}, \citenamefont {Nasu}, \citenamefont {Kato},\ and\ \citenamefont
  {Motome}}]{Yoshitake:2020}%
  \BibitemOpen
  \bibfield  {author} {\bibinfo {author} {\bibfnamefont {J.}~\bibnamefont
  {Yoshitake}}, \bibinfo {author} {\bibfnamefont {J.}~\bibnamefont {Nasu}},
  \bibinfo {author} {\bibfnamefont {Y.}~\bibnamefont {Kato}},\ and\ \bibinfo
  {author} {\bibfnamefont {Y.}~\bibnamefont {Motome}},\ }\href@noop {}
  {\bibfield  {journal} {\bibinfo  {journal} {Physical Review B}\ }\textbf
  {\bibinfo {volume} {101}},\ \bibinfo {pages} {100408} (\bibinfo {year}
  {2020})}\BibitemShut {NoStop}%
\bibitem [{\citenamefont {Gordon}\ \emph {et~al.}(2019)\citenamefont {Gordon},
  \citenamefont {Catuneanu}, \citenamefont {S{\o}rensen},\ and\ \citenamefont
  {Kee}}]{Gordon:2019}%
  \BibitemOpen
  \bibfield  {author} {\bibinfo {author} {\bibfnamefont {J.~S.}\ \bibnamefont
  {Gordon}}, \bibinfo {author} {\bibfnamefont {A.}~\bibnamefont {Catuneanu}},
  \bibinfo {author} {\bibfnamefont {E.~S.}\ \bibnamefont {S{\o}rensen}},\ and\
  \bibinfo {author} {\bibfnamefont {H.-Y.}\ \bibnamefont {Kee}},\ }\href@noop
  {} {\bibfield  {journal} {\bibinfo  {journal} {Nature communications}\
  }\textbf {\bibinfo {volume} {10}},\ \bibinfo {pages} {1} (\bibinfo {year}
  {2019})}\BibitemShut {NoStop}%
\bibitem [{\citenamefont {Gammel}\ \emph {et~al.}(1988)\citenamefont {Gammel},
  \citenamefont {Schneemeyer}, \citenamefont {Wasczak},\ and\ \citenamefont
  {Bishop}}]{Gammel1988}%
  \BibitemOpen
  \bibfield  {author} {\bibinfo {author} {\bibfnamefont {P.}~\bibnamefont
  {Gammel}}, \bibinfo {author} {\bibfnamefont {L.}~\bibnamefont {Schneemeyer}},
  \bibinfo {author} {\bibfnamefont {J.}~\bibnamefont {Wasczak}},\ and\ \bibinfo
  {author} {\bibfnamefont {D.}~\bibnamefont {Bishop}},\ }\href@noop {}
  {\bibfield  {journal} {\bibinfo  {journal} {Physical review letters}\
  }\textbf {\bibinfo {volume} {61}},\ \bibinfo {pages} {1666} (\bibinfo {year}
  {1988})}\BibitemShut {NoStop}%
\bibitem [{\citenamefont {Kleiman}\ \emph {et~al.}(1985)\citenamefont
  {Kleiman}, \citenamefont {Kaminsky}, \citenamefont {Reppy}, \citenamefont
  {Pindak},\ and\ \citenamefont {Bishop}}]{Bishop-RSI-1985}%
  \BibitemOpen
  \bibfield  {author} {\bibinfo {author} {\bibfnamefont {R.~N.}\ \bibnamefont
  {Kleiman}}, \bibinfo {author} {\bibfnamefont {G.~K.}\ \bibnamefont
  {Kaminsky}}, \bibinfo {author} {\bibfnamefont {J.~D.}\ \bibnamefont {Reppy}},
  \bibinfo {author} {\bibfnamefont {R.}~\bibnamefont {Pindak}},\ and\ \bibinfo
  {author} {\bibfnamefont {D.~J.}\ \bibnamefont {Bishop}},\ }\href
  {https://doi.org/10.1063/1.1138425} {\bibfield  {journal} {\bibinfo
  {journal} {{\it Review of Scientific Instruments }}\ }\textbf {\bibinfo
  {volume} {56}},\ \bibinfo {pages} {2088} (\bibinfo {year}
  {1985})}\BibitemShut {NoStop}%
\bibitem [{\citenamefont {Modic}\ \emph {et~al.}(2018)\citenamefont {Modic},
  \citenamefont {Bachmann}, \citenamefont {Ramshaw}, \citenamefont {Arnold},
  \citenamefont {Shirer}, \citenamefont {Estry}, \citenamefont {Betts},
  \citenamefont {Ghimire}, \citenamefont {Bauer}, \citenamefont {Schmidt},
  \citenamefont {Baenitz}, \citenamefont {Svanidze}, \citenamefont {McDonald},
  \citenamefont {Shekhter},\ and\ \citenamefont {Moll}}]{Modic2018}%
  \BibitemOpen
  \bibfield  {author} {\bibinfo {author} {\bibfnamefont {K.~A.}\ \bibnamefont
  {Modic}}, \bibinfo {author} {\bibfnamefont {M.~D.}\ \bibnamefont {Bachmann}},
  \bibinfo {author} {\bibfnamefont {B.~J.}\ \bibnamefont {Ramshaw}}, \bibinfo
  {author} {\bibfnamefont {F.}~\bibnamefont {Arnold}}, \bibinfo {author}
  {\bibfnamefont {K.~R.}\ \bibnamefont {Shirer}}, \bibinfo {author}
  {\bibfnamefont {A.}~\bibnamefont {Estry}}, \bibinfo {author} {\bibfnamefont
  {J.~B.}\ \bibnamefont {Betts}}, \bibinfo {author} {\bibfnamefont {N.~J.}\
  \bibnamefont {Ghimire}}, \bibinfo {author} {\bibfnamefont {E.~D.}\
  \bibnamefont {Bauer}}, \bibinfo {author} {\bibfnamefont {M.}~\bibnamefont
  {Schmidt}}, \bibinfo {author} {\bibfnamefont {M.}~\bibnamefont {Baenitz}},
  \bibinfo {author} {\bibfnamefont {E.}~\bibnamefont {Svanidze}}, \bibinfo
  {author} {\bibfnamefont {R.~D.}\ \bibnamefont {McDonald}}, \bibinfo {author}
  {\bibfnamefont {A.}~\bibnamefont {Shekhter}},\ and\ \bibinfo {author}
  {\bibfnamefont {P.~J.~W.}\ \bibnamefont {Moll}},\ }\href
  {https://doi.org/10.1038/s41467-018-06412-w} {\bibfield  {journal} {\bibinfo
  {journal} {{\it Nature Communications }}\ }\textbf {\bibinfo {volume} {9}},\
  \bibinfo {pages} {3975} (\bibinfo {year} {2018})}\BibitemShut {NoStop}%
\bibitem [{\citenamefont {Callen}(1985)}]{Callen1985}%
  \BibitemOpen
  \bibfield  {author} {\bibinfo {author} {\bibfnamefont {H.~B.}\ \bibnamefont
  {Callen}},\ }\href@noop {} {\emph {\bibinfo {title} {Thermodynamics and an
  introduction to thermostatics}}}\ (\bibinfo  {publisher} {John Wiley \&
  Sons},\ \bibinfo {year} {1985})\BibitemShut {NoStop}%
\bibitem [{\citenamefont {Lampen-Kelley}\ \emph {et~al.}(2018)\citenamefont
  {Lampen-Kelley}, \citenamefont {Janssen}, \citenamefont {Andrade},
  \citenamefont {Rachel}, \citenamefont {Yan}, \citenamefont {Balz},
  \citenamefont {Mandrus}, \citenamefont {Nagler},\ and\ \citenamefont
  {Vojta}}]{Lampen2018}%
  \BibitemOpen
  \bibfield  {author} {\bibinfo {author} {\bibfnamefont {P.}~\bibnamefont
  {Lampen-Kelley}}, \bibinfo {author} {\bibfnamefont {L.}~\bibnamefont
  {Janssen}}, \bibinfo {author} {\bibfnamefont {E.}~\bibnamefont {Andrade}},
  \bibinfo {author} {\bibfnamefont {S.}~\bibnamefont {Rachel}}, \bibinfo
  {author} {\bibfnamefont {J.-Q.}\ \bibnamefont {Yan}}, \bibinfo {author}
  {\bibfnamefont {C.}~\bibnamefont {Balz}}, \bibinfo {author} {\bibfnamefont
  {D.}~\bibnamefont {Mandrus}}, \bibinfo {author} {\bibfnamefont
  {S.}~\bibnamefont {Nagler}},\ and\ \bibinfo {author} {\bibfnamefont
  {M.}~\bibnamefont {Vojta}},\ }\href@noop {} {\bibfield  {journal} {\bibinfo
  {journal} {arXiv preprint arXiv:1807.06192}\ } (\bibinfo {year}
  {2018})}\BibitemShut {NoStop}%
\bibitem [{\citenamefont {Riedl}\ \emph {et~al.}(2019)\citenamefont {Riedl},
  \citenamefont {Li}, \citenamefont {Winter},\ and\ \citenamefont
  {Valent{\'\i}}}]{Riedl:2019}%
  \BibitemOpen
  \bibfield  {author} {\bibinfo {author} {\bibfnamefont {K.}~\bibnamefont
  {Riedl}}, \bibinfo {author} {\bibfnamefont {Y.}~\bibnamefont {Li}}, \bibinfo
  {author} {\bibfnamefont {S.~M.}\ \bibnamefont {Winter}},\ and\ \bibinfo
  {author} {\bibfnamefont {R.}~\bibnamefont {Valent{\'\i}}},\ }\href@noop {}
  {\bibfield  {journal} {\bibinfo  {journal} {Physical review letters}\
  }\textbf {\bibinfo {volume} {122}},\ \bibinfo {pages} {197202} (\bibinfo
  {year} {2019})}\BibitemShut {NoStop}%
\bibitem [{\citenamefont {Blundell}(2000)}]{Blundell}%
  \BibitemOpen
  \bibfield  {author} {\bibinfo {author} {\bibfnamefont {S.}~\bibnamefont
  {Blundell}},\ }\href@noop {} {\emph {\bibinfo {title} {Magnetism in Condensed
  Matter}}}\ (\bibinfo  {publisher} {Oxford Master Series in Physics},\
  \bibinfo {year} {2000})\BibitemShut {NoStop}%
\bibitem [{\citenamefont {Anderson}(1973)}]{Anderson:1973}%
  \BibitemOpen
  \bibfield  {author} {\bibinfo {author} {\bibfnamefont {P.~W.}\ \bibnamefont
  {Anderson}},\ }\href@noop {} {\bibfield  {journal} {\bibinfo  {journal}
  {Materials Research Bulletin}\ }\textbf {\bibinfo {volume} {8}},\ \bibinfo
  {pages} {153} (\bibinfo {year} {1973})}\BibitemShut {NoStop}%
\bibitem [{\citenamefont {Akiyama}\ \emph {et~al.}(2010)\citenamefont
  {Akiyama}, \citenamefont {de~Rooij}, \citenamefont {Staufer}, \citenamefont
  {Detterbeck}, \citenamefont {Braendlin}, \citenamefont {Waldmeier},\ and\
  \citenamefont {Scheidiger}}]{Akiyama2010}%
  \BibitemOpen
  \bibfield  {author} {\bibinfo {author} {\bibfnamefont {T.}~\bibnamefont
  {Akiyama}}, \bibinfo {author} {\bibfnamefont {N.~F.}\ \bibnamefont
  {de~Rooij}}, \bibinfo {author} {\bibfnamefont {U.}~\bibnamefont {Staufer}},
  \bibinfo {author} {\bibfnamefont {M.}~\bibnamefont {Detterbeck}}, \bibinfo
  {author} {\bibfnamefont {D.}~\bibnamefont {Braendlin}}, \bibinfo {author}
  {\bibfnamefont {S.}~\bibnamefont {Waldmeier}},\ and\ \bibinfo {author}
  {\bibfnamefont {M.}~\bibnamefont {Scheidiger}},\ }\href
  {https://doi.org/10.1063/1.3455219} {\bibfield  {journal} {\bibinfo
  {journal} {{\it Review of Scientific Instruments }}\ }\textbf {\bibinfo
  {volume} {81}},\ \bibinfo {pages} {063706} (\bibinfo {year}
  {2010})}\BibitemShut {NoStop}%
\bibitem [{cus()}]{custom-PLL}%
  \BibitemOpen
  \href@noop {} {}\bibinfo {note} {Custom software that implements phase-locked
  loop is available at
  \href{https://github.com/arkadyshekhter/frequencyshift}{https://github.com/arkadyshekhter/frequencyshift}.}\BibitemShut
  {Stop}%
\bibitem [{\citenamefont {Winter}\ \emph {et~al.}(2016)\citenamefont {Winter},
  \citenamefont {Li}, \citenamefont {Jeschke},\ and\ \citenamefont
  {Valent{\'\i}}}]{Winter:2016}%
  \BibitemOpen
  \bibfield  {author} {\bibinfo {author} {\bibfnamefont {S.~M.}\ \bibnamefont
  {Winter}}, \bibinfo {author} {\bibfnamefont {Y.}~\bibnamefont {Li}}, \bibinfo
  {author} {\bibfnamefont {H.~O.}\ \bibnamefont {Jeschke}},\ and\ \bibinfo
  {author} {\bibfnamefont {R.}~\bibnamefont {Valent{\'\i}}},\ }\href@noop {}
  {\bibfield  {journal} {\bibinfo  {journal} {Physical Review B}\ }\textbf
  {\bibinfo {volume} {93}},\ \bibinfo {pages} {214431} (\bibinfo {year}
  {2016})}\BibitemShut {NoStop}%
\end{thebibliography}

%

\end{document}